\documentclass[twocolumn, tighten, table,x11names]{aastex631}

\usepackage{longtable}
\usepackage[]{xcolor}
\definecolor{lightlightgray}{gray}{0.9} %

\DeclareRobustCommand{\okina}{%
  \raisebox{\dimexpr\fontcharht\font`A-\height}{%
    \scalebox{0.8}{`}%
  }%
}

\submitjournal{PSJ}

\shorttitle{}
\shortauthors{Deam et al.}
\graphicspath{{./}{figures/}}

\begin{document}

\title{A portrait throughout perihelion of the NH$_2$-rich interstellar comet 2I/Borisov}

\correspondingauthor{Michele Bannister}
\email{michele.bannister@canterbury.ac.nz}

\author[0000-0003-1955-628X]{Sophie E. Deam}
\affiliation{School of Physical and Chemical Sciences -- Te Kura Mat\={u}, University of Canterbury, Private Bag 4800, Christchurch 8140, New Zealand}
\affiliation{Space Science and Technology Centre, School of Earth and Planetary Sciences, Curtin University, Perth, Western Australia 6845, Australia}

\author[0000-0003-3257-4490]{Michele T. Bannister}
\affiliation{School of Physical and Chemical Sciences -- Te Kura Mat\={u}, University of Canterbury, Private Bag 4800, Christchurch 8140, New Zealand}

\author[0000-0002-9298-7484]{Cyrielle Opitom} 
\affiliation{Institute for Astronomy, University of Edinburgh, Royal Observatory, Edinburgh EH9 3HJ, UK}

\author[0000-0003-2781-6897]{Matthew M.~Knight}
\affiliation{Physics Department, United States Naval Academy, 572C Holloway Rd, Annapolis, MD 21402, USA}
\affiliation{University of Maryland, Department of Astronomy, College Park, MD 20742, USA}

\author[0000-0003-1724-2885]{Ryan Ridden-Harper}
\affiliation{School of Physical and Chemical Sciences -- Te Kura Mat\={u}, University of Canterbury, Private Bag 4800, Christchurch 8140, New Zealand}

\author[0000-0002-0726-6480]{Darryl Z. Seligman}
\affiliation{Department of Physics and Astronomy, Michigan State University, East Lansing, 48824, MI, USA}

\author[0000-0003-0250-9911]{Alan Fitzsimmons}
\affiliation{Astrophysics Research Centre, School of Mathematics and Physics, Queen's University Belfast, Belfast BT7 1NN, United Kingdom}

\author[0000-0003-2354-0766]{Aur{\'e}lie Guilbert-Lepoutre} 
\affiliation{Laboratoire de G{\'e}ologie de Lyon, LGL-TPE, UMR 5276 CNRS / Universit{\'e} de Lyon / Universit{\'e} Claude Bernard Lyon 1 / ENS Lyon, 69622 Villeurbanne, France}

\author[0000-0001-8923-488X]{Emmanuel Jehin}
\affiliation{STAR Institute, Universit\'e de Li\`ege, All\'ee du 6 ao\^ut, 19C, 4000 Li\`ege, Belgium}

\author{Laurent Jorda} 
\affiliation{Aix Marseille Univ, CNRS, LAM, Laboratoire d'Astrophysique de Marseille, Marseille, France}

\author[0000-0001-8617-2425]{Michael Marsset}
\affiliation{European Southern Observatory, Alonso de C\'ordova 3107, Santiago, Chile}

\author[0000-0001-9784-6886]{Youssef Moulane}
\affiliation{School of Applied and Engineering Physics, Mohammed VI Polytechnic University, Ben Guerir 43150, Morocco}

\author[0000-0001-6798-7805]{Philippe Rousselot} 
\affiliation{Université Marie et Louis Pasteur, CNRS, Institut UTINAM (UMR 6213), OSU THETA, F-25000 Besançon, France}

\author{Pierre Vernazza} 
\affiliation{Aix Marseille Univ, CNRS, LAM, Laboratoire d'Astrophysique de Marseille, Marseille, France}

\author{Bin Yang}
\affiliation{Instituto de Estudios Astrofísicos, Facultad de Ingeniería y Ciencias, Universidad Diego Portales, Av. Ej\'ercito Libertador 441, Santiago, 8370191, Chile}

\begin{abstract}
The interstellar comet 2I/Borisov is the first interstellar object where compositional characterisation was possible throughout its entire perihelion passage.
We report all 16 epochs of a comprehensive optical observation campaign with ESO VLT's integral field spectrograph MUSE, spanning 126 days from 2019 November 14 to 2020 March 19.
The spatial dust emission of 2I/Borisov was predominantly smooth, with no seasonal effect.
A jet-like feature was consistently visible.
The gas production morphology of its coma was also smooth and similar for C$_2$, NH$_2$, and CN: symmetric around the photocentre.
The production rates of these species gently declined into and beyond perihelion, until 2I's outburst and splitting event in early 2020 March. 
C$_2$, NH$_2$, and CN production rates all increased, with NH$_2$ being the most significant; the dust emission also slightly reddened. %
2I/Borisov is a carbon-depleted, relatively NH$_2$-rich comet when compared to those comets yet measured in the Solar System.

\end{abstract}


\section{Introduction} 
\label{sec:intro}

Interstellar objects (ISOs), interstellar interlopers \citep{Jewitt:2023} or sola lapidae \citep[`lonely rocks';][]{Portegies:2018} are small bodies that originate from outside the Solar System. 
The discovery of ISOs in the past eight years caused great excitement in the planetary community \citep{ISSITeam:2019,MoroMartin:2022,Jewitt:2023,Fitzsimmons:2023,Seligman2023}. 

Classically, ISOs are inferred to be remnants of the planetesimal formation process.  
Their physical properties could therefore help us constrain aspects of the disk chemistry and planetary architecture of their origin system \citep{Miotello:2023}. 
These small bodies are ejected into interstellar space, for instance by gravitational scattering from other bodies in the disk \citep{Raymond2018,Childs2022}, the stellar cluster \citep{Hands2019,Pfalzner2021}, or later post-main-sequence unbinding \citep[e.g.][]{Levine:2023}, and orbit in the Galaxy \citep{Forbes:2025}.
A tiny fraction of the vast galactic ISO population continuously pass through the Solar System's observable volume, where they can be identified and characterised with the same techniques as small bodies native to the Solar System.
ISOs provide an opportunity to directly sample other planetary systems, comparing the outcomes of planetesimal formation both in our Solar System and in distant systems.

Of the first two known macroscopic ISOs, 1I/\okina Oumuamua \citep{Williams17} and 2I/Borisov, discovered in August 2019\footnote{\textit{Minor Planet Electronic Circular} 2023-U162}, only the second had the detectable coma helpful for chemistry characterisation of internal material previously shielded from interstellar medium processing.
1I/\okina Oumuamua generated intense discussion, due to its lack of a detectable coma, but moderate radial (or anti-Solar) non-gravitational acceleration \citep{Micheli:2018}, and high-amplitude lightcurve, which implied an unusually high-aspect ratio 6:1 shape with an oblate geometry \citep{Mashchenko2019}.
In contrast, 2I/Borisov had an extended coma already visible at the time of detection and in precovery data \citep{Ye2020}.

Like Solar System comets, 2I underwent post-perihelion changes.
Two outbursts were detected between 2020 March 4-9 \citep{Drahus:2020} resulting in a total brightness increase of 2I by 0.7 magnitudes over the 5 days. 
The March outbursts coincided with a splitting of the nucleus, confirmed when the fragment became resolvable by the Hubble Space Telescope (HST) on March 30, 2020 \citep{Jewitt:2020}. 
The splitting involved the ejection of a fragment with a mass around 0.01\% of the nucleus.

2I/Borisov exhibited many traits with strong similarities to Solar System comets. 
Color measurements showed that the dust was redder than the Sun \citep{Guzik:2020, Hui:2020, de-Leon:2020, Kareta:2020, Fitzsimmons:2019, Mazzotta-Epifani:2021, Yang:2020, Yang:2021}, and the coma was dust-rich with properties similar to those of 67P/Cheryumov-Gerasimenko \citep{Busarev:2021}. 
The coma's structure at discovery showed featured similar to jets, with two enhancements that were generally perpendicular to the Solar direction \citep{Mazzotta-Epifani:2021, Bolin:2020, Manzini:2020}.
Radio observations of 2I using occultation and interplanetary scintillation techniques found small-scale density structures in the range of $\sim$10-70~km in the plasma tail \citep{Manoharan:2022}. 
Spectroscopic measurements showed further similarity to Solar System comets, with an early detection of CN \citep{Fitzsimmons:2019} and later C$_2$ emission at optical wavelengths. 
C$_2$ was depleted relative to CN, like the `carbon-depleted' group \citep{Fitzsimmons:2019,Opitom:2019-borisov,Bannister:2020,Kareta:2020,Lin:2020,Aravind:2021}.
Higher-resolution spectroscopy revealed gaseous nickel \citep{Guzik:2021}, and a comparable NiI/FeI abundance ratio and ammonia ortho-to-para ratio \citep{Opitom:2021}.

However, 2I/Borisov was unusual relative to almost all Solar System comets in one respect: it was extremely rich in CO relative to HCN and water \citep{Bodewits:2020,Cordiner:2020}.  
The only comet with a coma richer in CO than 2I to date is C/2016R2 \citep{Biver:2018}.
This high CO measurement was corroborated by the high ratio of the green-to-red doublet intensity of the three [OI] forbidden oxygen lines measured by \cite{Opitom:2021}, which has been suggested to indicate a significant contribution of CO or CO$_2$ \citep{McKay:2024} to the outgassing compared to H$_2$O.

These properties have influenced discussion of the environment in which 2I could have formed and its subsequent evolution. 
Based on its high CO abundance, \citet{Cordiner:2020} and \citet{Bodewits:2020} argued that 2I probably originated outside the CO line of its home planetary disk. 
There is also the possibility that 2I is an outer fragment of a larger-differentiated object, as proposed for C/2016R2 \citep{Biver:2018}.
\citet{Bagnulo:2021} argued the homogeneous, high degree of polarisation of 2I in comparison to Solar System comets indicates it had not passed close to any star since its formation \citep[see also][]{Halder2023}.
As post-ejection stellar encounters are unlikely \citep{Forbes:2025}, one can infer that any thermal alteration of 2I, which is minimal, occurred in its host system and is strongly connected to its dynamical history.

The high velocities of ISOs \citep{Hopkins:2025} mean that the period during which their spectroscopic characterisation is possible is brief, predicted to be a few hundred days if found when inbound \citep{Dorsey:2025}.
For 1I, detected post-perihelion, only a few weeks of observation were possible for most facilities before it grew too faint. 
Extensive multi-month datasets of major facility characterisation could not be acquired with anything other than HST \citep[e.g. see][]{Jewitt:2023}.
In contrast, 2I/Borisov was the first ISO detected pre-perihelion.
Upon discovery at a heliocentric distance of 2.98~au, its brightness was sufficient for characterisation with 8-m class ground-based telescopes.

We conducted an extensive observing campaign of 2I/Borisov through its perihelion passage with ESO VLT's Multi Unit Spectroscopic Explorer (MUSE) from 2019 November 14 onward.
Due to the COVID-19 pandemic and associated closure of telescopes to safeguard the health of all involved, the program ended earlier than the limiting magnitude of MUSE, on 2020 March 19 --- the last observation made of this interstellar comet with a major ground-based facility\footnote{The last ground-based observation of 2I reported to the Minor Planet Center was astrometry from New Zealand's Mt John Observatory in Takap\={o}, 2020 April 28; when New Zealand returned to Covid-19 Alert Level 3.}. 
The initial results of the gas production rates of C$_2$, NH$_2$, \& CN and dust continuum slope for the 2019 November 14, 15 \& 26 observations were reported in \citet{Bannister:2020}. 

Here, we report on the MUSE dataset from the entire 126-day span of the observing campaign. 
With a new star-subtraction package for integral field unit (IFU) spectrographs, built specifically to account for the dense stellar fields that 2I moved across \citep[\texttt{starkiller}:][]{Ridden-Harper2025} (\S~\ref{sec:making_maps}), we study the spatial distribution of 2I's outgassing with maps of the solar reflectance gradients (\S~\ref{sec:dustcolour}) and dust (\S~\ref{sec:dust}).
We assess spatial gas emission (\S~\ref{sec:gasmaps}) and computed gas production rates through the perihelion passage of 2I and beyond (\S~\ref{sec:production}).
Finally, we consider the variability of 2I and its aspects of unusual composition relative to Solar System comets (\S~\ref{sec:discussion}).

\section{Observations}
\label{sec:obs}

\begin{figure*}[ht!]
\centering
\includegraphics[width=0.75\textwidth]{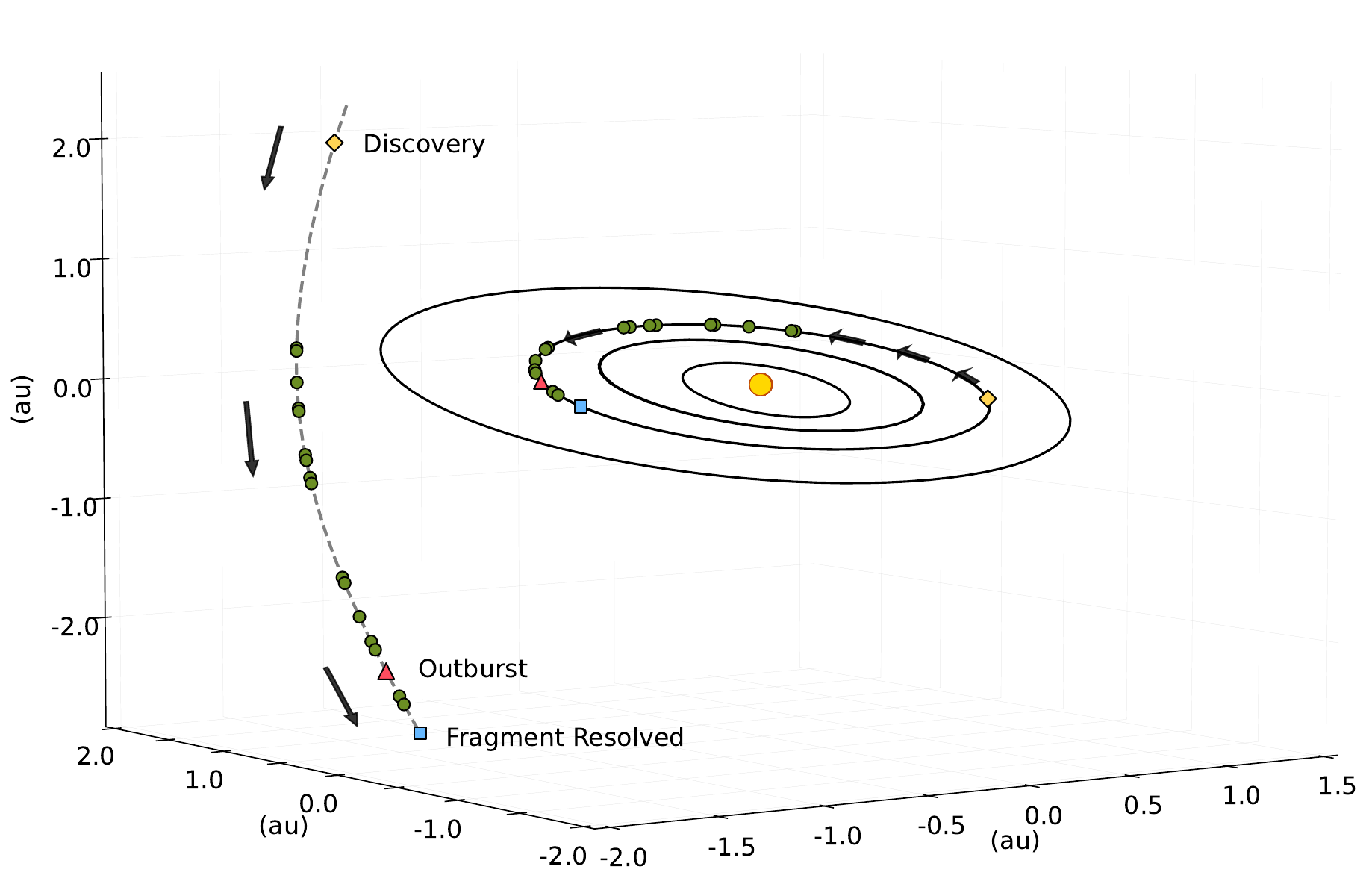}
\caption{Relative positions of 2I/Borisov and other Solar System bodies during the VLT/MUSE observing campaign, shown in International Celestial Reference Frame (ICRF) coordinates. 
The trajectory of 2I follows the dashed line in the direction of the arrows. 
Green circles indicate the positions of 2I and Earth at each of the 16 observing epochs (Table~\ref{observations_full_condensed}). 
Coloured symbols are additional events identified independent of our data: yellow diamonds represent 2I and Earth's positions at discovery on 2019 August 30, orange triangles the outburst of 2020 March 4-9 \citep{Drahus:2020}, and blue squares when the fragment was resolved by HST on 2020 March 30 \citep{Jewitt:2020}. 
The black lines outward from the Sun (yellow circle) are the orbits of Mercury, Venus, Earth, and Mars.} 
\label{2Itrajectory}
\end{figure*}

2I/Borisov was observed with ESO VLT's MUSE from 2019 November 14 to 2020 March 19, under the ESO Director’s Discretionary programs 103.2033.001--003 and Normal program 105.2086.002. 
The campaign acquired 16 epochs, which are summarised in Table~\ref{observations_full_condensed}. 
Figure~\ref{2Itrajectory} illustrates 2I's trajectory through the Solar System and the extensive portion of its journey covered by MUSE observations. 
All observations, including reduced data products, are publicly available\footnote{The DOI link to the data archive at CADC will be added following manuscript review. Access to the archive is available on request if desired by the reviewer.}, with analysis scripts also available\footnote{\url{https://github.com/sed79/super-duper-comet}}.

\startlongtable
\begin{deluxetable*}{ccccccccccc}

\tablecaption{VLT/MUSE observations of 2I/Borisov, with geometric relationships between 2I, Earth, and the Sun. %
}
\tablehead{\colhead{Observation Date} & \colhead{N$^1$} & \colhead{Airmass$^2$} & \colhead{Seeing$^3$} & \colhead{$r_h$$^4$} & \colhead{$\Delta$$^5$} & \colhead{Elongation$^6$} & \colhead{$\alpha$$^7$} & \colhead{$\theta_{-\odot}$$^8$} & \colhead{$\theta_{-V}$$^9$} & \colhead{$l^{10}$}\\ 
\colhead{(UTC)}  & \colhead{} & \colhead{} & \colhead{($"$)} & \colhead{($au$)} & \colhead{($au$)} & \colhead{($\mathrm{{}^{\circ}}$)} & \colhead{($\mathrm{{}^{\circ}}$)} & \colhead{($\mathrm{{}^{\circ}}$)} & \colhead{($\mathrm{{}^{\circ}}$)} & \colhead{($km$)}} 
\startdata
2019-Nov-14 08:03-08:48 & 4\tablenotemark{a} & 2.17 & 0.8 & 2.08 & 2.23 & 68.25 & 26.26 & 289.29 & 330.98 & 1617 \\ 
2019-Nov-15 08:02-08:53 & 4\tablenotemark{a} & 2.13 & 0.87 & 2.07 & 2.22 & 68.60 & 26.40 & 289.25 & 330.94 & 1608 \\ 
2019-Nov-26 07:17-08:03 & 4 & 2.16 & 0.43 & 2.02 & 2.09 & 72.41 & 27.69 & 289.11 & 330.02 & 1517 \\ 
2019-Dec-05 07:09-07:55 & 4 & 1.82 & 0.93 & 2.01 & 2.02 & 75.40 & 28.36 & 289.50 & 328.65 & 1462 \\ 
2019-Dec-06 07:37-08:25 & 4\tablenotemark{b} & 1.55 & 1.04 & 2.01 & 2.01 & 75.73 & 28.42 & 289.57 & 328.45 & 1457 \\ 
2019-Dec-21 07:56-08:13 & 4\tablenotemark{a} & 1.21 & 0.49 & 2.03 & 1.94 & 80.39 & 28.60 & 291.52 & 324.50 & 1410 \\ 
2019-Dec-23 05:50-06:42 & 4 & 1.98 & 0.39 & 2.03 & 1.94 & 80.96 & 28.55 & 291.89 & 323.84 & 1407 \\ 
2019-Dec-29 05:18-06:09 & 4 & 2.12 & 0.66 & 2.06 & 1.94 & 82.74 & 28.30 & 293.22 & 321.54 & 1405 \\ 
2019-Dec-31 05:36-06:22 & 4 & 1.83 & 0.41 & 2.07 & 1.94 & 83.33 & 28.18 & 293.74 & 320.68 & 1406 \\ 
2020-Feb-02 04:17-05:03 & 4 & 1.66 & 0.45 & 2.35 & 2.08 & 93.18 & 24.71 & 307.01 & 302.04 & 1512 \\ 
2020-Feb-04 03:18-04:05 & 4\tablenotemark{c} & 2.01 & 0.55 & 2.38 & 2.10 & 93.81 & 24.44 & 308.12 & 300.82 & 1522 \\ 
2020-Feb-16 04:07-05:04 & 5\tablenotemark{d} & 1.53 & 0.6 & 2.53 & 2.20 & 97.83 & 22.77 & 315.86 & 293.87 & 1593 \\
2020-Feb-25 03:28-04:11 & 4 & 1.57 & 0.81 & 2.65 & 2.28 & 101.03 & 21.49 & 322.76 & 289.78 & 1653 \\ 
2020-Feb-28 04:02-04:48 & 4\tablenotemark{e} & 1.45 & 0.72 & 2.70 & 2.31 & 102.14 & 21.06 & 325.32 & 288.69 & 1674 \\ 
2020-Mar-16 04:41-05:28 & 4\tablenotemark{f} & 1.33 & 1.23 & 2.96 & 2.48 & 108.59 & 18.61 & 341.85 & 285.55 & 1801 \\ 
2020-Mar-19 05:23-06:14 & 4\tablenotemark{e} & 1.31 & 0.61 & 3.00 & 2.52 & 109.74 & 18.19 & 345.14 & 285.50 & 1826 \\ 
\enddata

\tablenotetext{a}{Two frames discarded due to high twilight background}
\tablenotetext{b}{One frame discarded due to high twilight background}
\tablenotetext{c}{Three frames omitted from production rate analysis due to 2I passing in front of a star}
\tablenotetext{d}{One frame discarded due to thin cloud conditions}
\tablenotetext{e}{One frame discarded and one frame omitted from production rate analysis due to 2I passing in front of a star}
\tablenotetext{f}{Two frames omitted from production rate analysis due to 2I passing in front of a star}
\tablecomments{1: Number of frames, 2: Starting airmass, 3: FWHM seeing value at the beginning of the observation measured by the ASM-DIMM telescope, 4: Heliocentric distance, 5: Geocentric distance, 6: Solar elongation angle, 7: Solar phase angle, 8: Position angle of the anti-Solar direction, 9: Position angle of the negative velocity vector, 10: Physical size of 1" projected to the distance of 2I in km.}
\label{observations_full_condensed}
\end{deluxetable*}

MUSE operates on the Nasmyth platform of the 8.2 m diameter optical telescope UT4 at ESO's Paranal Observatory in Chile.
This instrument has a large field of view ($1{\arcmin}{\times}1{\arcmin}$) covered by 24 individual integral field units (IFUs), ideal for imaging extended sources \citep{Bacon:2010}. 
The nominal MUSE wavelength range spans 4830~\AA{}--9300~\AA{}, with a spatial sampling of $0.2{\arcsec}{\times}0.2{\arcsec}$ and spectral resolution of 1.25~\AA. 
This covers the emission bands of the Swan C$_2$($\Delta v=0$, $\Delta v=1$) system, several amino radical NH$_2$ bands, and the red CN A$^{2}\Pi$-X$^{2}\Sigma^{+}$ system. 
MUSE was used in the wide-field mode without adaptive optics for all of the observations reported in this paper. 
At each epoch, 2I/Borisov was imaged for 4 $\times$ 600~s exposures. 
Interjecting sky exposures were obtained after the first and third images, intended to be used for clear sky measurements, but ended up not being used in the analysis (see~\S~\ref{sec:reduction}). 
Sky exposures were integrated for 180 s during epochs 2019 November 14 through to 2020 February 4, and were reduced to 120 s for 2020 February 16 onward, because the field surrounding 2I became increasingly crowded as the comet passed into the plane of the Milky Way. 
Small dithering and 90$^\circ$ rotation were applied between each of the 4 target exposures to allow for the removal of detector signatures. 
Standard stars were observed each night to flux-calibrate the data (Table \ref{tab:standardstars} in Appendix~\ref{sec:AppendixObservingInfo}).

The 2019 November 14 images are off-centered within the field of view due to ephemeris uncertainties\footnote{The ephemerides from IMCCE differed from those from JPL Horizons. Subsequent observations only used Horizons.} that were subsequently resolved. 
Observations on all nights except 2020 February 16 were made in clear or photometric conditions (Table~\ref{observations_full_condensed}). 
The 3rd exposure obtained that night was affected by clouds, so an additional exposure was acquired and the affected exposure discarded. 
However, the extra image does not have the same rotation orientation as the image it was replacing, so the detector signals were not as efficiently removed after co-adding. 
The strong background from morning twilight and the close proximity of background stars prevented us from using a number of exposures (Appendix~\ref{sec:AppendixObservingInfo}, Table~\ref{observations_full}).

\section{Data Reduction and Analysis}
\label{sec:reduction}

\subsection{Initial Data Reduction}

The MUSE data reduction pipeline \citep{Weilbacher:2020} performed dark subtraction, flat fielding, telluric correction, flux calibration, and datacube reconstruction. 
Datacubes were sky subtracted in two different ways: using in-frame sky measurements and offset sky exposures. %
For the first sets of observations on November 14 and 15, we used the in-frame sky measurements for subtraction, as the scientific exposures of 2I were taken deep in twilight and suffered from a significant and varying sky background. 
In theory, using in-frame sky measurement could lead to an over-subtraction of the sky if the object occupies a significant portion of the field of view. 
For these observations, we performed tests with the two different sky techniques, comparing the production rates derived, and did not find any sign of over-subtraction when using in-frame sky measurements. 
We thus decided to use that technique for the entire dataset. 
The output cubes were made with wavelength calibration to arc line wavelengths in air (as opposed to vacuum) as is default for the MUSE reduction pipeline.

Telluric absorption was dealt with in two different ways. For creating maps of the gas and dust, we used the pipeline telluric correction, based on observations of the standard stars. 
Since the standard stars were not observed at the same time or at the same airmass as the object, this method can under- or over-correct the telluric absorption. 
This is not a problem when looking at gas maps, but is more critical when deriving CN production rates, since that region of the spectrum is significantly affected by telluric absorptions. 
For the purpose of deriving production rates we therefore skipped the pipeline telluric correction and performed it directly on the extracted spectra using the Molecfit software \citep{Smette2015}.

\subsection{Map Creation}
\label{sec:making_maps}

Maps of 2I's continuum, solar reflectance, and emission of the radicals CN, C$_2$, and NH$_2$ were created using data-cubes obtained in photometric or clear conditions (with the exception of 2020 February 16; see Table \ref{observations_full_condensed}). 
2I's passage in front of the galactic plane during the post-perihelion observations meant up to $\sim$ 400 stars of 21st magnitude in GAIA-G and brighter were present in a single exposure. %

Due to the increasingly numerous stars in the FOV, the observations on and after 2019 December 31 were processed with a tool named \texttt{starkiller}\footnote{Commit da0b884: \href{https://github.com/CheerfulUser/starkiller/tree/da0b884d1fde0175d412f605b2a42cb914d6cca8}{https://github.com/CheerfulUser/starkiller}},  specifically developed to remove streaked sources in integral field unit datacubes \citep{Ridden-Harper2025}. 
Each datacube was passed to \texttt{starkiller} individually before extracting or combining maps.  
The removal of streaks overlapping with 2I's coma allowed an extra 4 datacubes to be included in the map creation where stellar contamination had initially led to discarding them. 
Only two exposures were discarded because the stars behind 2I could not be sufficiently removed.
\texttt{starkiller} also improved the image quality derived from a further 27 datacubes. 

\begin{figure*}

    \includegraphics[width=1\textwidth]{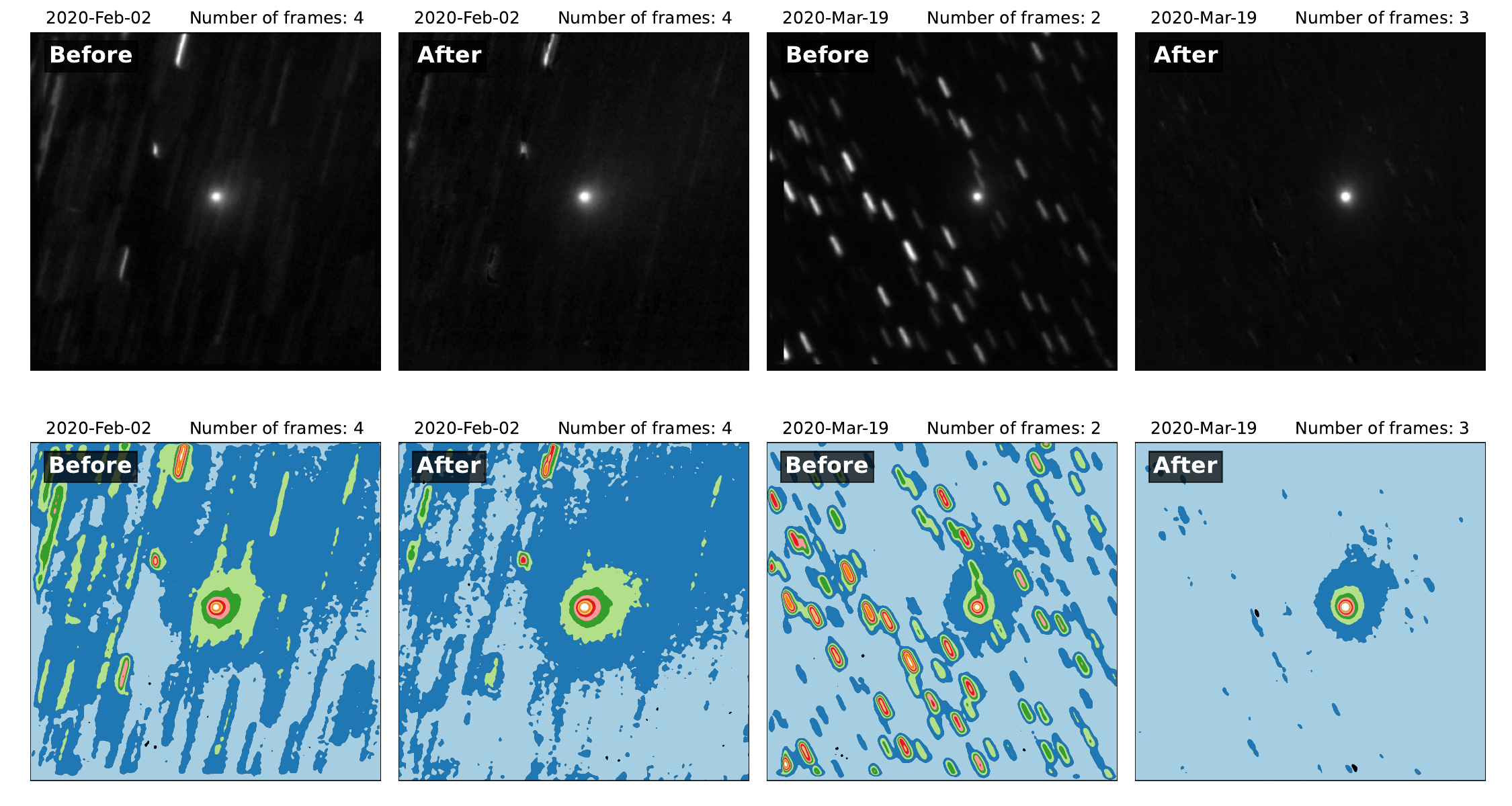}
    \caption{Images of the dust emission (7080\AA{}--7120\AA{}) from 2I/Borisov before and after the application of the \texttt{starkiller} pipeline. 2I was within 10 degrees of the Galactic plane on 2020-Feb-02 and 2020-Mar-19, and the FOV for each exposure contained between 80 and 136 stars brighter than 21st magnitude in GAIA-G. More significant improvement occurred for the 2020-Mar-19 data: the number of usable exposures (without a star directly behind 2I) increased from 2 to 3 due to \texttt{starkilller}, allowing for improved median co-adding and the removal of more stars. Both rows show the maps with a square root stretch, and the contour levels in the bottom row illustrates the reduction of star flux to below the signal from the coma.}
\label{fig:beforeafterstarkiller}
\end{figure*}

A comparison of maps created from datacubes processed with and without \texttt{starkiller} are reproduced from \citet{Ridden-Harper2025} in Figure \ref{fig:beforeafterstarkiller}.
The quality of the star subtraction is wavelength-dependent, due to pipeline processes that keep the subtraction as unbiased as possible. 
This means that the subtraction quality varies between the maps produced here. 
It was more effective in the CN range than the dust images in Figure \ref{fig:beforeafterstarkiller}.
\texttt{starkiller} was most successful when it prevented a star-contaminated exposure from being discarded to increase the number of frames being co-added together and the overall signal-to-noise ratio of the final data product, such as 2020 February 04 and 2020 March 19. 

After the star removal, the extreme pixel values (upper and lower 5e$^{-6}$ percentiles) caused by cosmic rays and other sources of noise were masked. 
The comet's optocentre (centre of brightness) was determined with the \texttt{photutils} 2D Gaussian centroid algorithm.
To spatially align each data cube and place 2I consistently in the centre, a uniform shift was applied to each image slice using 3rd-order spline interpolation with \texttt{MUSE Python Data Analysis Framework (MPDAF)}\citep{MPDAFsoftware}.

\subsubsection{Dust and Colour Maps}
\label{sec:methoddustcolour}

Dust maps were extracted by collapsing aligned datacubes in the spectral dimension over 7080~\AA{}-7120~\AA{} (Table \ref{wavelength ranges}), a wavelength range that contains few gas or sky emission lines \citep{Farnham2000}. 
Inspecting the maps, we do not see any sign of rotation by 2I. 
We stacked the observations from a single night obtained over 1 hour, and additionally stacked together 2019 December 5-6 which were only 24h apart, to increase the signal-to-noise ratio of the data when investigating coma structure. 
The physical scale in the image changed between December 5 and 6 as the distance from the Earth to 2I decreased. In other words, although the images were aligned at the coma centre, coma features were misaligned by up to 200~km at the image edge, which we did not adjust for before co-adding between dates.
We justified this decision by verifying that the relevant features --- or lack thereof --- were evident in the exposures from individual nights, i.e. the co-adding does not affect our interpretation. 

In order to create maps of the dust colour, each datacube was also divided by a reference Solar spectrum \citep{Meftah:2018}, which was interpolated using \texttt{scipy.interpolate} to match the datacube's spectral sampling. 
Three wavelength ranges of width 40~\AA{} were extracted from regions of each datacube without gas emission, and then averaged along the spectral dimension to produce three maps centred around the wavelengths 5280~\AA{}, 7080~\AA{}, \& 8600~\AA{}. 
The 7080~\AA{} map was used to study the dust morphology, while the other two were used to produce colour maps.

The normalised reflectivity gradient was calculated for each pixel using the definition of \citet{Jewitt:1986},

\begin{equation}
    S'(\lambda_1, \lambda_2)=(dS/d\lambda)/S_{mean}
\end{equation}

where $\lambda_1, \lambda_2$ are the wavelengths 5280~\AA{} \& 8600~\AA{}, $S$ is the solar divided spectrum such that $dS/d\lambda$ is the gradient of $S$ over 5280~\AA{}--8600~\AA{}, and $S_{mean}$ is the mean within the wavelength range 5280~\AA{}--8600~\AA{}, for which we substitute the average flux around 7080~\AA{} because it avoids emission bands and best represents the continuum.
Where $S_{mean}=0$, $S'$ was simply set to zero to avoid division by $S_{mean}$. 
Colour maps were stacked in the same way as dust maps.
The median value and root-mean-square error of $S'$ within a 5,000~km diameter aperture were calculated from the maps to report a single $S'$ value for each night.
To assess any changes in $S'$ after the reported March outburst, a linear regression line was fitted to the data preceding the outburst that was weighted by the measurement uncertainties. 
The post-outburst data was then compared with the 95\% or 2$\sigma$ prediction confidence interval for the corresponding dates to conclude whether there was a statistically significant change.%

\subsubsection{Gas Emission Maps}

To isolate the spectrum of 2I's gas emission, the light scattered by the dust in the coma was removed from the datacubes by subtracting a spectrum of comet 67P/Churyumov–Gerasimenko (67P), previously obtained by MUSE \citep{Guilbert-Lepoutre:2016,Opitom2020}, which does not contain any gas emission. 
First, to account for differences in the dust colour caused by phase angle and composition (expected between 2I and 67P), the spectrum of 2I was divided by the 67P spectrum at each spatial location of a datacube. 
A second-order polynomial was then fitted to the resulting spectrum over regions containing no gas emission, to produce a model polynomial. 
This was used to weight and slope correct the spectrum of 67P to match the gradient of 2I.
The 67P spectrum was then normalised and subtracted from each 2I spectrum to produce dust-subtracted datacubes. 
As the same subtraction technique was applied to all spatial locations within the data cube, the remaining stars in the FOV after removal by \texttt{starkiller} also had a spectrum of 67P subtracted from them. 
This led to over-subtraction or under-subtraction, since these stellar spectra have different colours and features to 67P, meaning that the stars in the gas maps appear as either dark or light streaks.
Due to the proximity of the CN(1-0) emission at 9109~\AA{} to MUSE's upper limiting wavelength at 9300~\AA{}, there existed a `dip feature' in the continuum of each spectrum that did not exist in the spectrum of 67P. 
The spectra in each datacube were further manually flattened over the region 8800~\AA{}--9300~\AA{} by applying a 7th order polynomial to a coadded spectrum with a well-defined `dip feature' to be used as a model continuum for scaling and subtraction from each spatial pixel's spectrum, creating a flat continuum under the CN emission.

Gas maps were extracted by collapsing the datacubes in the spectral dimension over the wavelength ranges outlined in Table \ref{wavelength ranges}. 
All regions extracted covering NH$_2$ emission were combined into a single image. 
Gas maps from different datacubes were stacked following the same procedure as for the dust.

\begin{deluxetable}{cccc}
\tablewidth{\textwidth}
\tablecaption{The electronic transitions and the corresponding wavelength ranges used to extract maps from MUSE datacubes.}
\tablehead{\colhead{Species} & \colhead{System} & \colhead{Transition} &\colhead{Wavelengths} \\ \colhead{-} & \colhead{-} &  \colhead{-} &\colhead{(\AA{})}}
\label{wavelength ranges}
\startdata
        Dust & - & - & 7080 - 7120 \\
        C$_2$ &  $\mathrm{d^3\Pi_g-A^3\Pi_u}$ & ($\Delta \nu=0$) & 5075 - 5171 \\
        NH$_2$ & $\mathrm{\tilde{A}^{2}A_{1}-\tilde{X}^2B_1}$ & (0,9,0)-(0,0,0) & 5973 - 5982  \\
        NH$_2$ & $\mathrm{\tilde{A}^{2}A_{1}-\tilde{X}^2B_1}$ & (0,9,0)-(0,0,0)+[O$^{1}$D] & 5991 - 6001 \\
        NH$_2$ & $\mathrm{\tilde{A}^{2}A_{1}-\tilde{X}^2B_1}$ & (0,9,0)-(0,0,0) & 6016 - 6026 \\
        NH$_2$ & $\mathrm{\tilde{A}^{2}A_{1}-\tilde{X}^2B_1}$ & (0,8,0)-(0,0,0) & 6330 - 6340 \\
        CN & A$^{2}\Pi$-X$^{2}\Sigma^{+}$ & (1-0) & 9135 - 9210 \\
\enddata
\tablecomments{Each wavelength range for NH$_2$ were combined into a single map.}
\end{deluxetable}

\subsection{Image Enhancement}
The technique of subtracting the azimuthal median was used to investigate the non-uniformity of the coma around the optocentre of 2I. 
Areas of the coma with enhancement values significantly above and below zero indicate potential morphological features, whereas an image with enhancement values close to zero indicates a uniform coma. 
For more examples of the technique, we direct the reader to \citet{Knight2017-67P}.

\subsection{Gas Production Rates}
To derive production rates, we extracted spectra over the same physical aperture of 5,000~km for each datacube. 
This aperture size is rather small compared to what is usually used to measure gas production rates, but was chosen to avoid stars contaminating the extracted spectra in the 2020 observations, when 2I was crossing a crowded field. 
Note that \texttt{starkiller} was not applied to the data cubes used in gas production rate measurements: reducing the aperture radius was generally sufficient to avoid stellar contamination in these measurements, and on rare occasions when it was not, we note it in Table \ref{observations_full}.

\begin{figure*}[ht!]
\centering
\includegraphics[width=1.0\textwidth]{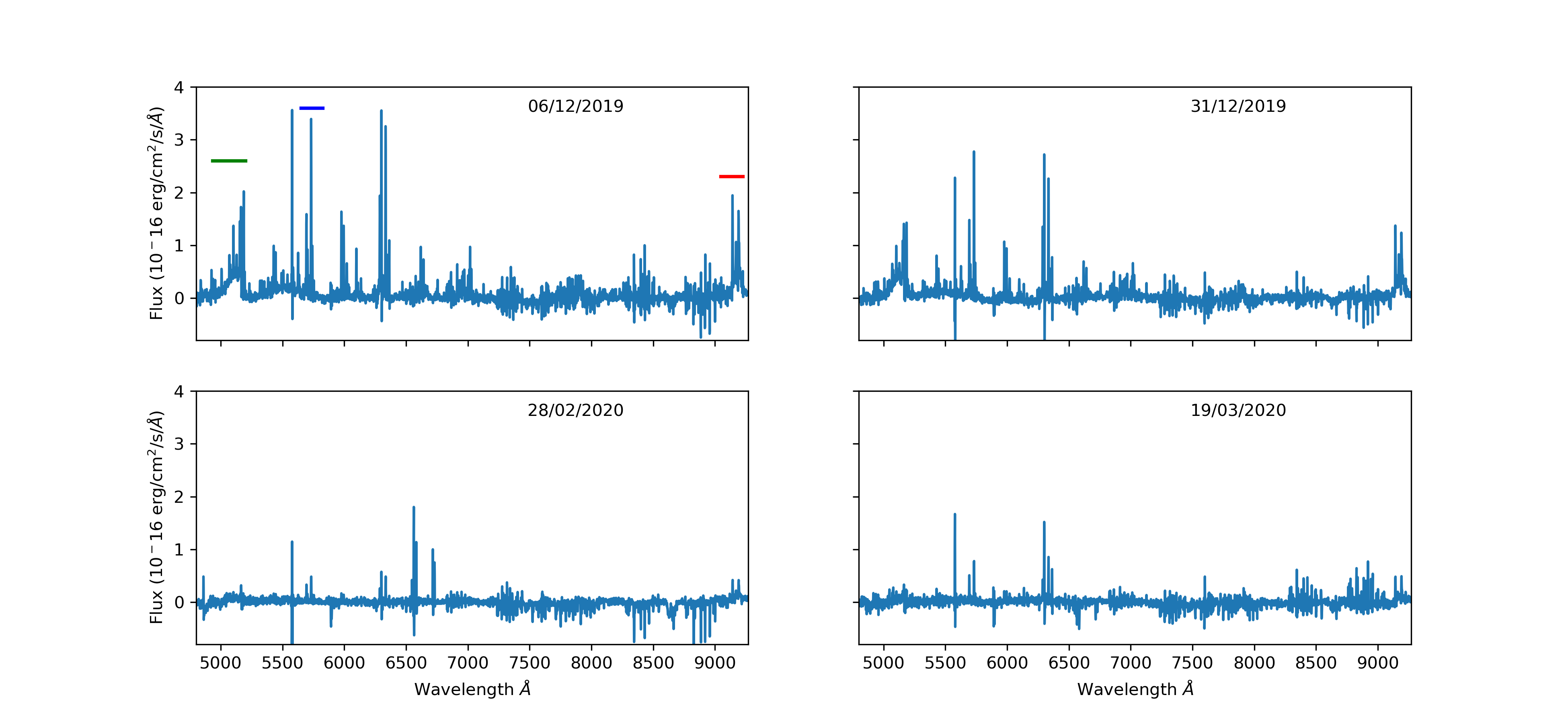}
\caption{Dust subtracted spectra of 2I at four different dates, extracted within a 5,000~km radius. The C$_2$, NH$_2$, and CN bands used to measure production rates are indicated by the green, blue, and red line respectively.} 
\label{2Ispectra}
\end{figure*}

The extracted spectra for each epoch were then averaged to increase the SNR. 
The telluric signatures were corrected using the Molecfit software \citep{Smette2015} and the continuum contribution was subtracted. 
For continuum subtraction, we used a spectrum of comet 67P obtained from MUSE data \citep[containing only the dust signature;][]{Guilbert-Lepoutre:2016,Opitom2020}, whose slope was adjusted to match the slope of 2I in each observation, similar to what was done to produce the gas maps. 
Examples of dust-subtracted spectra are shown in Figure \ref{2Ispectra}.

In the resulting spectra, emissions from the C$_2$ $\delta \nu=0$ and $\delta \nu=0-1$ bands, the NH$_2$ A–X (0,10,0)–(0,0,0), (0,9,0)–(0,0,0), (0,8,0)–(0,0,0), and (0,7,0)–(0,0,0) bands, and the red CN $(1-0)$ band around 9140 ~\AA\ are clearly visible. 
As 2I/Borisov moved away from the Sun in 2020, fainter bands progressively became undetectable in our observations. 
However, the C$_2$ $\delta \nu=0$, NH$_2$ A–X (0,10,0)–(0,0,0), and CN $(1-0)$ bands remained detectable throughout our campaign and were used to measure gas production rates. 

To derive the gas production rates, we measured the flux in each band for each epoch, using the continuum subtracted spectra. 
For CN, we simply integrated the spectra over the 9100–-9270~\AA\ range. 
The CN $(1-0)$ band contains flux at longer wavelengths, but we limited the integration to that range to avoid telluric and second order contamination at longer wavelengths. 
We then multiplied the flux by a factor 1.4 to account for the missing flux \citep{Fink:2009}. 
For the C$_2$ $\delta \nu=0$ band, we integrated the flux over the 4800--5170~\AA\ range. 
For NH$_2$, we adjusted a set of gaussians to reproduce the peaks of the (0,10,0)–(0,0,0), and then integrated them to obtain the total flux in the band in addition of simply integrating the flux over the wavelength range of the band. Both techniques resulted in similar fluxes being measured. 

\begin{deluxetable}{lccc}
\tabletypesize{\footnotesize}
\tablecolumns{4}
\tablecaption{Molecular fluorescence efficiencies (g-factors - $g$) and Haser model scale-lengths for parent and daughter species $l_p$, $l_d$ at $r_H=1$ au, used with $v=0.5$ km/s}
\label{tab:model_parms}
\tablehead{
\colhead{Band}
& \colhead{$g$}
& \colhead{$l_p$}
& \colhead{$l_d$} \\
\colhead{}
& \colhead{(ergs s$^{-1}$mol$^{-1}$)} 
& \colhead{(km)}  
& \colhead{(km)}
}
\startdata
C$_2(\Delta v=0)$ & $4.5\times10^{-13}$ & $2.2\times10^{4}$ & $6.6\times10^4$ \\ %
NH$_2$(0,10,0) & $9.2\times10^{-15}$ & $4.1\times10^{3}$ & $6.2\times10^{4}$ \\ 
CN(1-0) & $9.1\times10^{-14}$ & $1.3\times10^{4}$ & $2.1\times10^{5}$ \\ 
\enddata
\tablecomments{$g$-factors for C$_2$ are from \cite{Ahearn:1995}, for CN from \cite{Shinnaka2017}, and for NH$_2$ from \cite{Tegler:1989}. $g$-factors are assumed to scale as $r_{H}^{-2}$ and scalelengths  as $r_{H}^{2}$. Scalelengths are from \citet{Ahearn:1995} for C$_2$ and CN, and from \cite{Cochran:2012} for NH$_2$.
}
\end{deluxetable}

In parallel, we used the Planetary Spectrum Generator (PSG) \citep{Villanueva2018} to retrieve synthetic spectra that match the relative intensities of the bands observed for 2I. 
The synthetic spectra matched well the shape and intensity of the bands we observed, without noise produced by the sky and continuum subtraction. 
However, while C$_2$ production rates could be retrieved and matched closely those measured with the technique described above, production rates provided by the PSG retrieval were inconsistent with the production rates derived using the method described above for CN, potentially because of different models and model parameters used, and did not contain NH$_2$. We thus used the synthetic spectra to measure the flux in the CN and C$_2$ as for our observed spectra.

We converted the measured fluxes to column densities using the fluorescence efficiencies given in Table \ref{tab:model_parms} and used a Haser model \citep{Haser:1957} to derive gas production rates. 
We used a velocity of 0.5 km/s and the scalelengths listed in Table \ref{tab:model_parms}. 
We computed gas production rates using both the fluxes measured in our observed spectra and in our modelled spectra, which were consistent with each other. 
The production rates reported in Table \ref{tab:productionrates} are based on modelled spectra for CN and C$_2$, and the set of gaussians for NH$_2$, with the uncertainties estimated from the variation between the production rates derived from observed and modelled spectra. 
This uncertainty reflects the dominant contribution of the continuum subtraction and telluric correction in the gas production rate measurement. 

To test for any significant changes in the production rates and their ratios from the splitting of the nucleus, we fitted a trend to the post-perihelion data obtained prior to the March outburst, with the same method as for the dust colour described in \S~\ref{sec:methoddustcolour}.

\begin{deluxetable}{lllllll}
\tablecaption{Fluxes and production rates of the C$_2$, NH$_2$, \& CN gas emission of 2I/Borisov.}
\tablehead{
\colhead{Date} & 
\multicolumn{3}{c}{Flux ($10^{-15}$erg/s/cm$^2$)} & 
\multicolumn{3}{c}{Production Rate ($10^{24}$ mol/s)} \\
\colhead{(UTC)} &
\colhead{C$_2$} &
\colhead{NH$_2$} &
\colhead{CN} &
\colhead{$Q$(C$_2$)} &
\colhead{$Q$(NH$_2$)} &
\colhead{$Q$(CN)}
}
\startdata
2019-Nov-14  & 5.75 & 2.86 & 6.59 & 0.72$\pm$0.2 & 4.9$\pm$0.8 & 2.5$\pm$0.2 \\
2019-Nov-26  & 6.34 & 3.44 & 7.43 & 0.65$\pm$0.3 & 4.8$\pm$0.8 & 2.3$\pm$0.2 \\
2019-Dec-05  & 6.89 & 3.33 & 7.81 & 0.64$\pm$0.3 & 4.2$\pm$0.8 & 2.2$\pm$0.2 \\
2019-Dec-06  & 7.09 & 3.47 & 7.64 & 0.65$\pm$0.2 & 4.4$\pm$0.5 & 2.1$\pm$0.2 \\
2019-Dec-21    & 5.94 & 2.84 & 7.29 & 0.53$\pm$0.2 & 3.4$\pm$0.5 & 2.0$\pm$0.2 \\
2019-Dec-23    & 5.75 & 2.88 & 6.34 & 0.51$\pm$0.2 & 3.5$\pm$0.7 & 1.7$\pm$0.2 \\
2019-Dec-29    & 5.21 & 2.78 & - & 0.48$\pm$0.2 & 3.5$\pm$0.4 & - \\
2019-Dec-31    & 5.23 & 2.57 & 5.70 & 0.49$\pm$0.2 & 3.3$\pm$0.5 & 1.6$\pm$0.2 \\
2020-Feb-02    & 1.59 & 1.01 & 2.81 & 0.27$\pm$0.2 & 2.3$\pm$0.1 & 1.5$\pm$0.2 \\
2020-Feb-04    & 1.79 & 1.29 & 2.85 & 0.32$\pm$0.2 & 3.0$\pm$0.1 & 1.6$\pm$0.2 \\
2020-Feb-16    & 0.87 & 0.37 & 2.35 & 0.21$\pm$0.2 & 1.2$\pm$0.1 & 1.7$\pm$1.0 \\
2020-Feb-25    & 0.92 & 0.53 & 2.04 & 0.28$\pm$0.2 & 2.1$\pm$1.0 & 1.9$\pm$0.7 \\
2020-Feb-28    & 0.92 & 0.56 & 1.59 & 0.31$\pm$0.2 & 2.4$\pm$1.0 & 1.6$\pm$0.7 \\
2020-Mar-16    & 1.12 & 0.64 & 1.75 & 0.60$\pm$0.5 & 4.4$\pm$1.0 & 2.9$\pm$1.5 \\
2020-Mar-19    & 0.73 & 0.73 & 1.58 & 0.43$\pm$0.3 & 5.4$\pm$1.0 & 2.8$\pm$1.5 \\
\enddata
\tablecomments{Exposures from 2019-Nov-14 and 2019-Nov-15 were combined before the production rates were calculated. Observations from 2019-Dec-29 had additional telluric contamination over the CN wavelengths that were difficult to remove, so we do not report values for this date.}
\label{tab:productionrates}
\end{deluxetable}

\section{Results}
\label{sec:results}

\subsection{Dust Colour}
\label{sec:dustcolour}

\begin{figure*}
\includegraphics[width=1\textwidth]{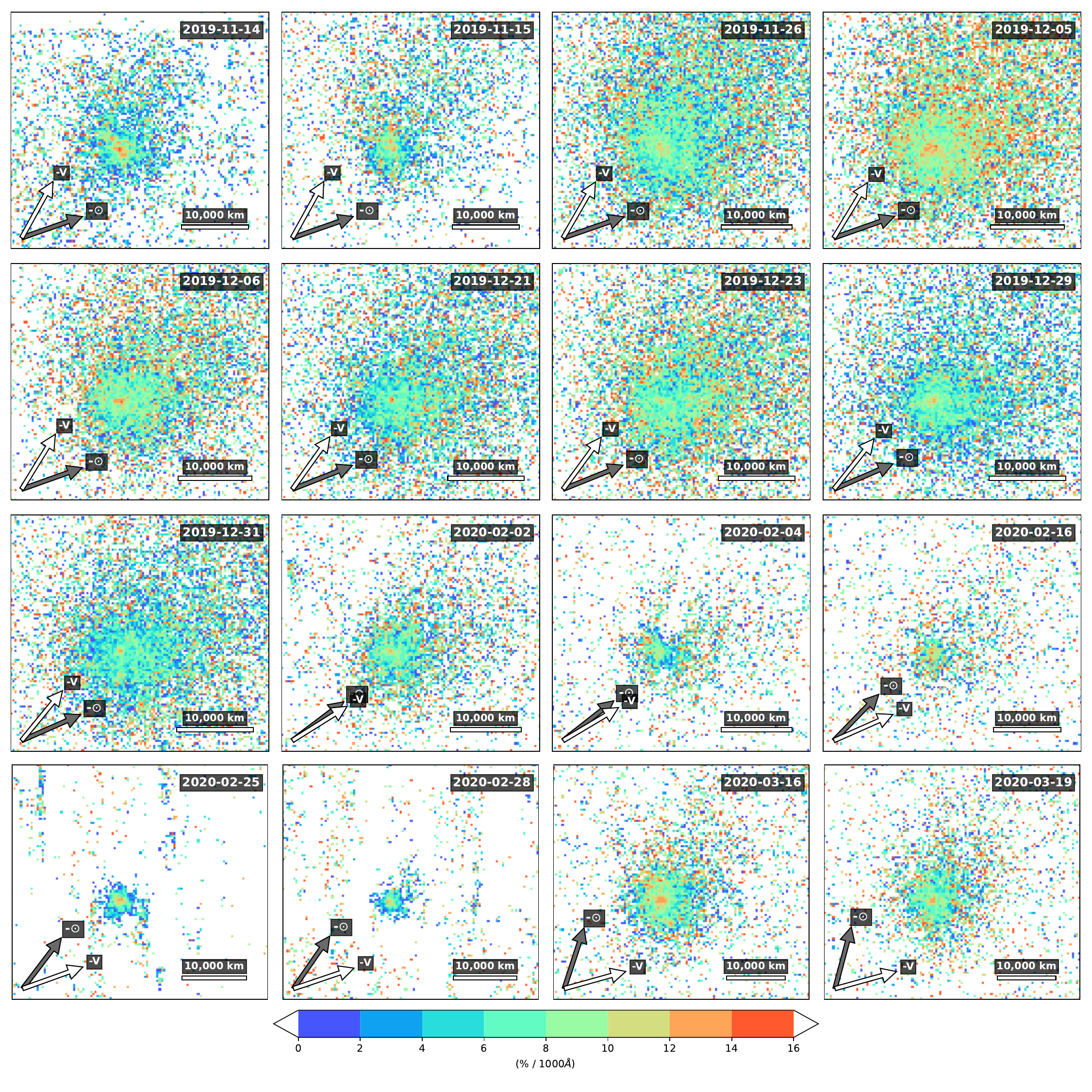}
\caption{Distribution of normalised solar reflectance gradients for 2I/Borisov. The very inner coma has a steeper gradient, between 10\%-14\% in comparison to the outer coma, within 4\% and 10\%. The extreme values in white represent the sky and background star streaks. North is up, east is to the left, and the anti-Solar (-$\odot$) and negative velocity (\textbf{-V}) directions are shown with the respective arrows.} 
\label{dustcolour}
\end{figure*}

The normalised solar reflectance gradient maps $S'$ within 5280~\AA{}--8600~\AA{} are shown in Figure \ref{dustcolour}. 2I's colour ranges from 4~\%/1000\AA{} to 10~\%/1000\AA{} and does not vary significantly through time over the MUSE observations.
The dust appears redder in the centre and bluer farther out in the coma. 
This has been seen previously, e.g. for C/2016 ER61 \citep{Opitom:2019-C2016ER61}, and is consistent with particle size sorting where larger (`redder') particles occupy the central part of the coma, and smaller (`bluer') particles occupy the outer regions of the coma. 
In most maps, we see a significantly redder central part, which could point at a condensed inner coma made of larger particles. 
The sky and background stars are plotted as white, either due to extreme slopes for stars or division by small sky fluxes around 7080~\AA{}. 
The noise produced by the division of low flux around 7080~\AA{} was not mitigated or masked for risk of altering the boundary between 2I and the sky.  

\begin{figure}[ht!]
\includegraphics[width=\columnwidth]{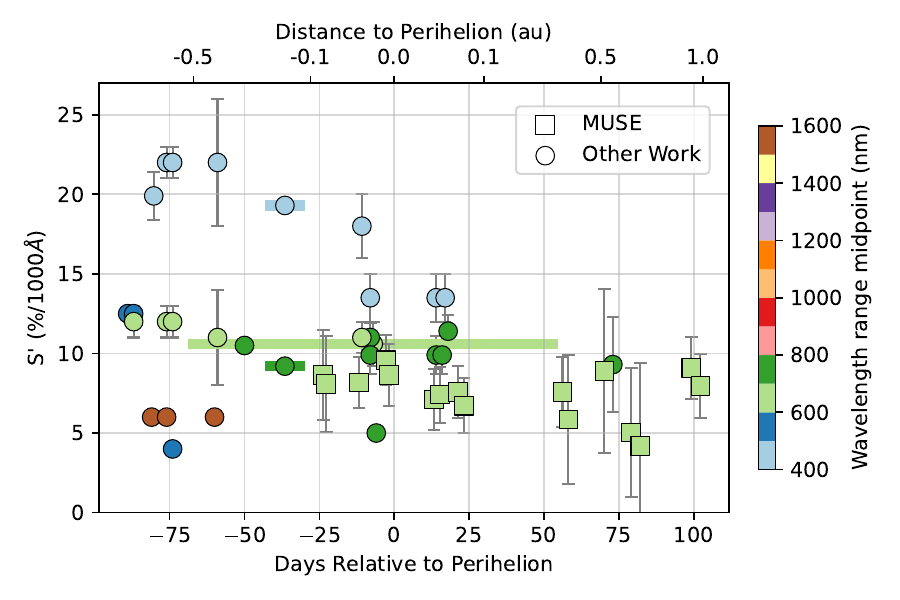}
\caption{$S'$ of 2I from both MUSE (squares) (5,000~km diameter aperture) and reported literature values (circles) relative to perihelion. Any reported value calculated from multiple observations has a horizontal line behind it spanning the range of dates. This data is also listed in Table \ref{S' table}, Appendix \ref{sec:AppendixDataforFigures}, along with the wavelength range over which $S'$ was measured. 
} 
\label{S' by Date}
\end{figure}

The dust colour over the 5,000~km diameter aperture extracted from MUSE data is consistent with other photometric methods, summarised in Table \ref{S' table} and illustrated in Figure \ref{S' by Date}. 
$S'$ is lower than other measurements taken in the visible, however the range 5280~\AA{}--8600~\AA{} used here extends further into the near IR. 
This is consistent with the concept of lower solar reflectance gradient at longer wavelengths of dust \citep{Jewitt:1986}, but makes it challenging to compare between data sets measured in different wavelength ranges.

\citet{Mazzotta-Epifani:2021} measured the reddening of 2I between the Bessel R and I filters over multiple annuli at increasing distances from the optocenter ($2{\times}10^{3}$~km through to $1.3{\times}10^{4}$~km), and find a decreasing slope i.e. reddening from
their inner to outer annuli for 2019 December 02. 
While most of the change of colour we observe with 2I is close to the optocenter and within \citet{Mazzotta-Epifani:2021}'s inner-most annulus radius of 2x10$^{3}$~km (one fifth of the scale bar in Figure \ref{dustcolour}), we also see the trend of decreasing spectral slope out to $1.3{\times}10^{4}$~km.

As mentioned above, we do not see a substantial change in dust colour with heliocentric distance or time. There is only a slight downward trend  of the dust colour over time (highlighted further in Figure \ref{fig:prodrateschange}). This is in contrast with previous reports from \citet{Busarev:2021} who found that the reflectance varied over time periods of a few days. They attributed that change to an increasing contamination from gas emission in their filters, which is not our case since we selected bands containing only continuum signal for our colour analysis.

\subsection{Dust Maps}
\label{sec:dust}

\begin{figure*}
    \includegraphics[width=1\textwidth]{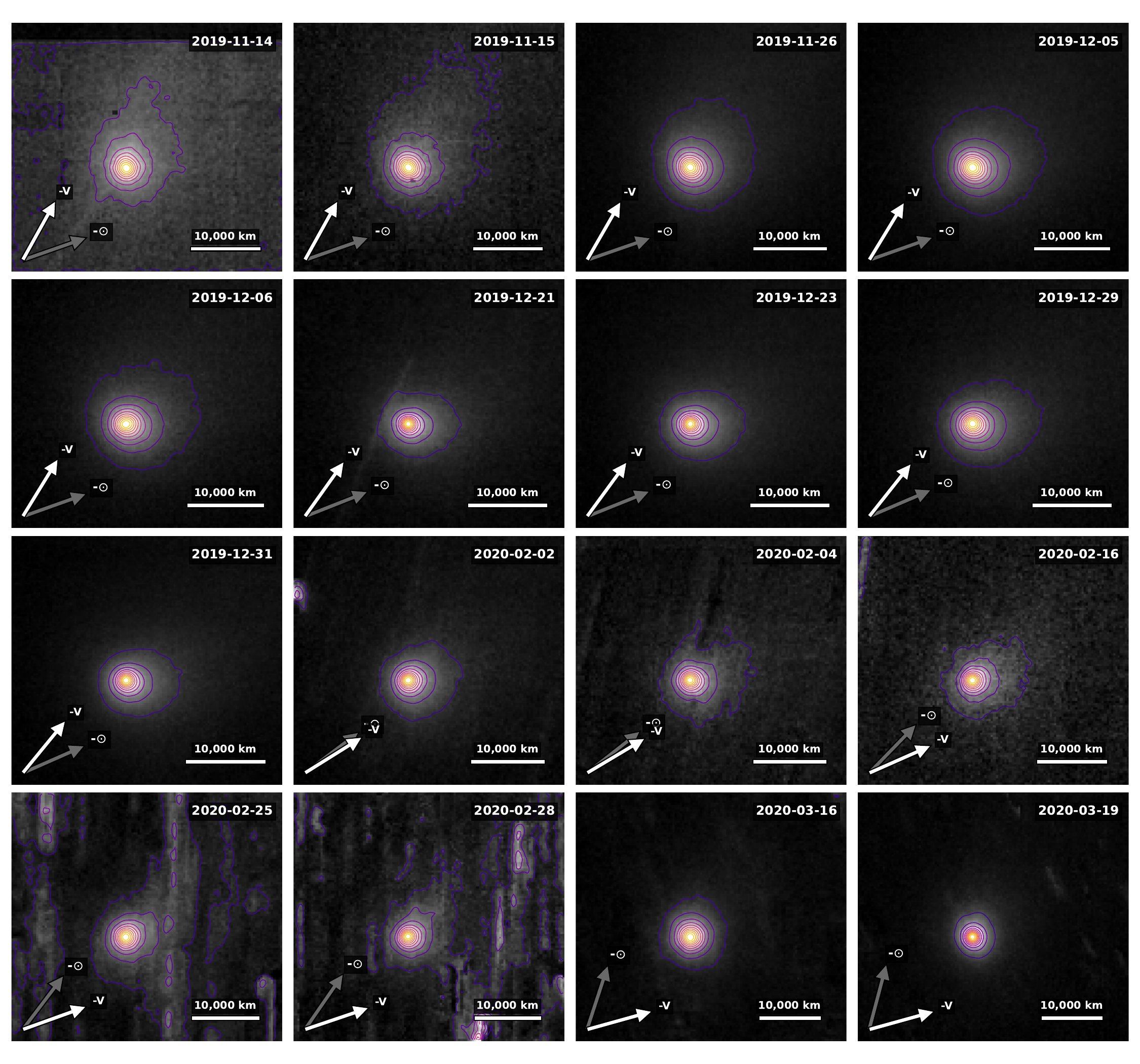}
    \caption{Dust emission maps spanning 2019 November 14 until 2020 March 19 displayed with a linear stretch and the same orientation as Fig. \ref{dustcolour}. 
    Streak effects are due to background stars (minimised by the application of \texttt{starkiller}).
    The contours (also fitted to the background stars and the edge of the field of view) at 9\% intervals show the elongation of the dust in the coma in relation to the anti-Solar (-$\odot$) and negative velocity (\textbf{-V}) directions.}
    \label{dustmapgrid}
\end{figure*}

\begin{figure*}
    \includegraphics[width=1\textwidth]{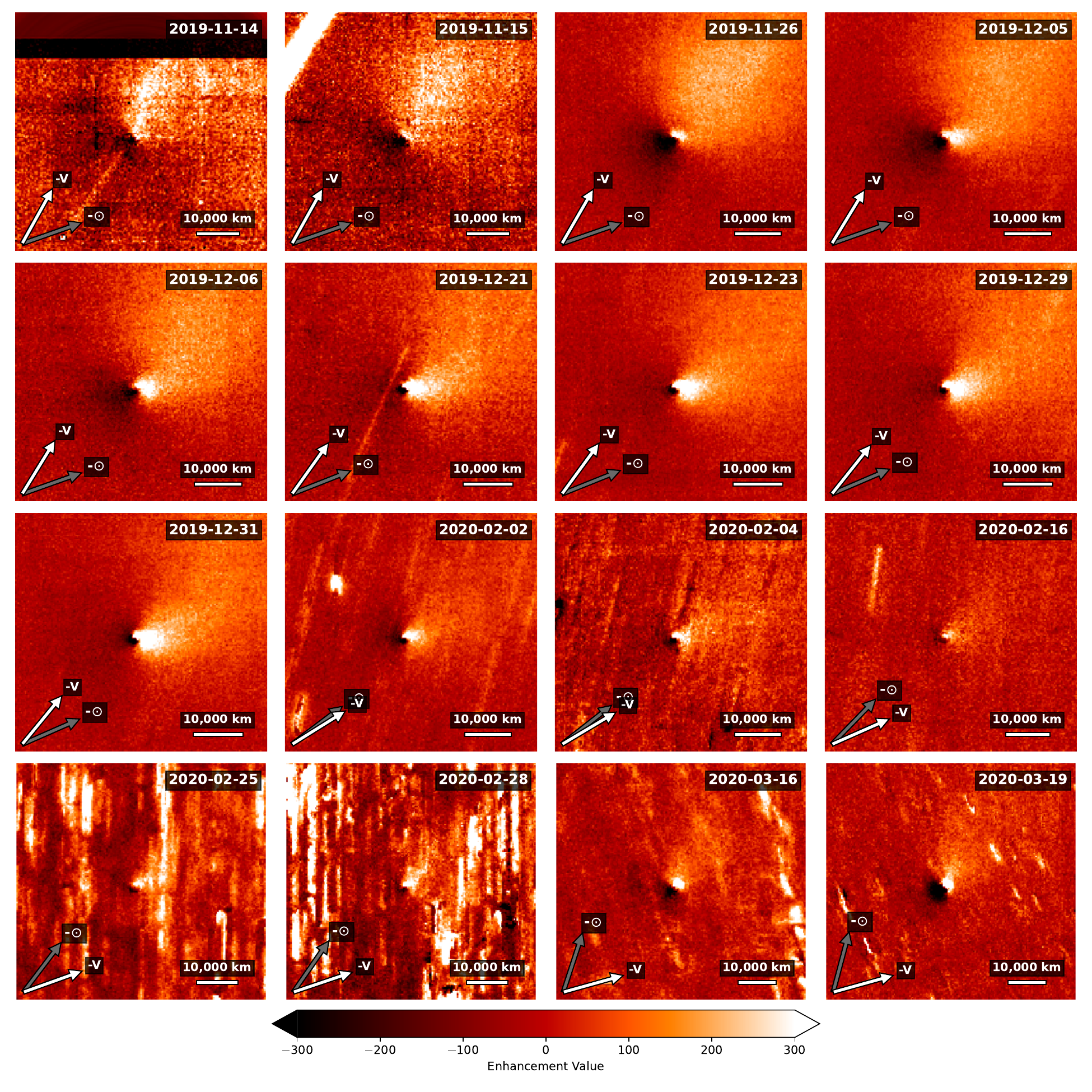}
    \caption{Dust maps from Figure \ref{dustmapgrid} enhanced by subtraction of azimuthal median with the same orientation as Fig. \ref{dustcolour}.
    Streak effects are due to background stars (minimised by the application of \texttt{starkiller}).
    There is a consistent jet-like feature towards the North-West indicated by higher enhancement values.
    The dark bands on 2019 November 14 are the edge of the image.}
    \label{dustmapenhancedgrid}
\end{figure*}

Figure \ref{dustmapgrid} presents the original dust maps, while Figure \ref{dustmapenhancedgrid} shows the same maps after image enhancement. 
The nights have varying levels of noise depending on the number of exposures co-added and the observing conditions. 
Several features can be distinguished in the dust maps. 
On November 14, 15, 26, and December 5 and 6, there is a clear elongation in Fig.~\ref{dustmapgrid} towards the North-West, corresponding to a broad feature in the enhanced maps of Figure \ref{dustmapenhancedgrid}.
Such asymmetry is also seen in \citet{Kim:2020}.
In Fig.\ref{dustmapenhancedgrid}, we see a jet-like feature towards the West, curving towards the North-West at larger nucleocentric distances. 
This becomes distinctly visible for all dates from November 26 onward.
The post-\texttt{starkiller} data permit observations with dense stellar crowding in February to be assessed.
We do not see any significant evolution of the dust morphology between the end of November 2019 and March 2020, i.e. through perihelion. 
\citet{Mazzotta-Epifani:2021}, \citet{Bolin:2020}, and \citet{Manzini:2020} all report observations of 2I in December 2019 and January 2020 (only January 2020 for \citet{Manzini:2020}). 
Their observations are consistent with what we see in the MUSE data: they report a jet-like feature towards the West. 
Since some of the observations are from HST --- with a much better spatial resolution than our MUSE data and processed with different techniques --- it would seem to confirm that the feature we observe towards the North-West is real and not an artifact of the enhancement technique we applied. 
These authors also report a smaller and fainter jet-like feature towards the East/North-East; we do not see this in our data.
Our dust maps (Fig.~\ref{dustmapgrid} and \ref{dustmapenhancedgrid}) do not resolve the March splitting event, but are otherwise consistent with the HST observations by \citet{Jewitt:2020} in the broadband F350LP filter, which saw general anti-Solar elongation in 2I's coma on 2020 March 23.

\subsection{Gas Maps}
\label{sec:gasmaps}

\begin{figure*}[ht!]
\centering
 \includegraphics[width=\textwidth]{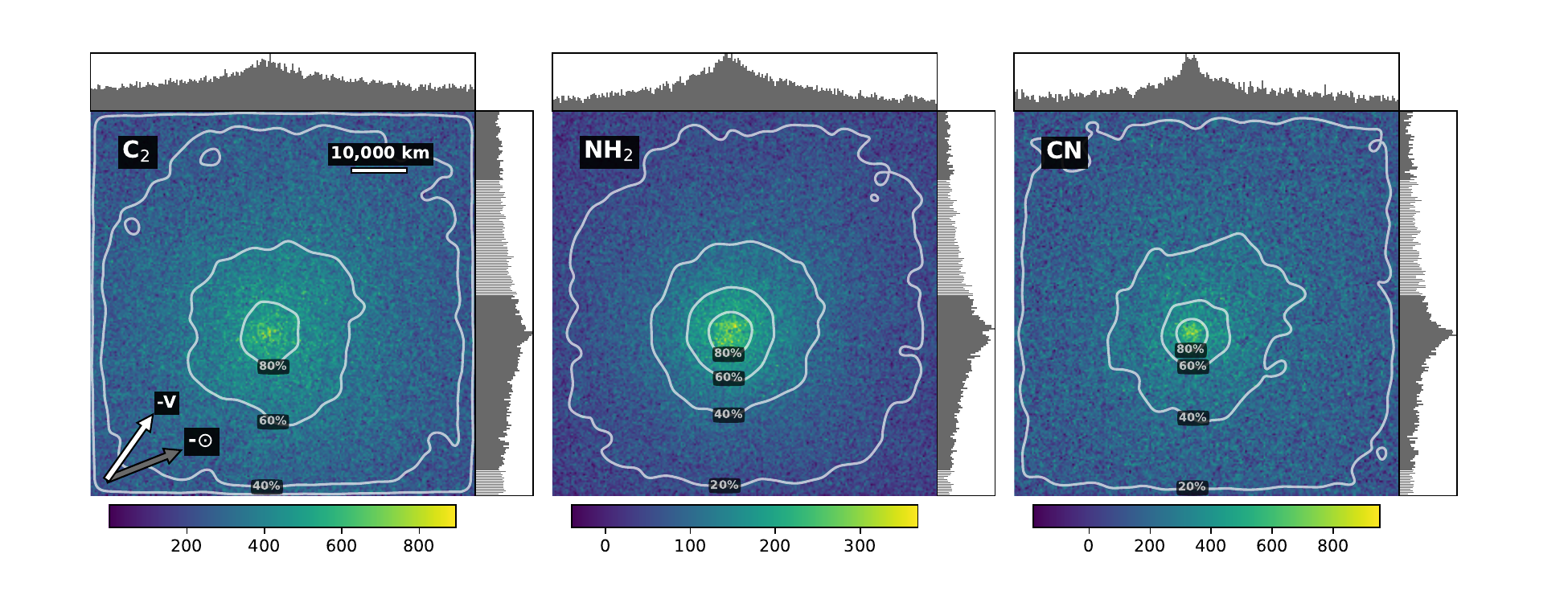}%
\caption{The relative spatial distribution of the C$_2$, NH$_2$, and CN gas using maps coadded from 2019 December 5 and 6 (7 $\times$ 600~s exposures). The histogram above/beside each map shows the sum of flux from a horizontal/vertical strip 10 pixels wide through the comet centre. All images are displayed with a linear stretch over a zmax scale and the same orientation as Fig. \ref{dustcolour}. The contours are 20\% flux intervals of the gas maps after smoothing by a 5 standard deviation Gaussian convolution. The 20\% flux line of C$_2$ is not labelled to avoid crowding. All images are 48 arcseconds ($\sim$70,000~km at the distance of 2I) wide. 
} 
\label{gasmapsdecember}
\end{figure*}

The spatial distribution of C$_2$, NH$_2$, and CN are presented in Figure~\ref{gasmapsdecember} for observations on 2019 December 5 and 6. 
This figure reveals a very similar morphology for C$_2$, NH$_2$, and CN. 
Their distribution is roughly symmetrical around the photocenter, with a hint of extension towards the North-West for CN. 
The maps for all dates are in Appendix~\ref{sec:Appendix:GasEmissionMaps}.
The comet is only detected in our C$_2$ and NH$_2$ maps in 2019 (with the exception of NH$_2$ in one post-outburst epoch), while CN is detected through 2019--2020. 
Taking into account the relatively low SNR of the maps, we do not see any significant changes of the gas morphology over the apparition. 
There appears to be a feature of CN emission to the North-West within 2,000~km of the optocentre of 2I on 2020 March 19; however, it was only visible in one of the three contributing data cubes.
We conclude it is contamination from background stars in the same position that was not suppressed during the median co-adding of exposures. 
The SNR was insufficient to reveal any morphological features in the gas maps under image enhancement (see Figure~\ref{EnhancedCNmapgrid}).

\subsection{Production Rates}
\label{sec:production}

\begin{figure*}
    \includegraphics[width=1\textwidth]{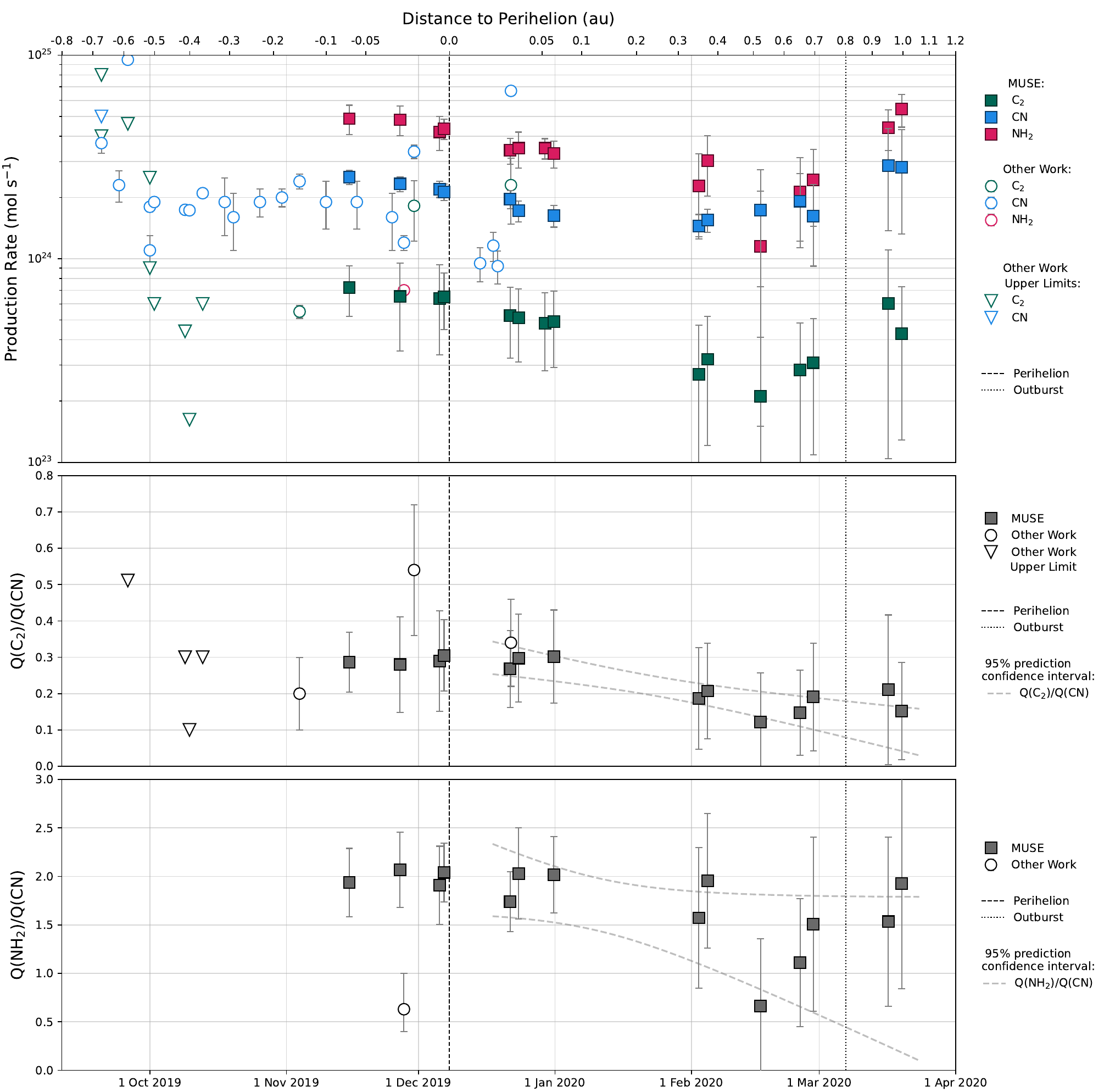}
    \caption{The C$_2$, NH$_2$, CN production rates (upper figure), $Q$(C$_2$)/$Q$(CN) ratio (middle figure), and $Q$(NH$_2$)/$Q$(CN) ratio (lower figure) for 2I from 20 September 2019 to 19 March 2020. Our measurements are shown next to other sources:  \cite{Fitzsimmons:2019, de-Leon:2019, Kareta:2020, Opitom:2019-borisov, Lin:2020, Bannister:2020, Aravind:2021, Cordiner:2020, Prodan:2024}. Note that the measurements of  $Q$(C$_2$)/$Q$(CN)$<$1 on 2019 September 20 \citep{Fitzsimmons:2019} and $Q$(C$_2$)/$Q$(CN)$<$2.3 on 2019 October 1 \citep{Kareta:2020} are omitted. The 95\% confidence interval windows for the production rate ratios are shown as light gray dashed lines, and are predicted for post-outburst dates by extrapolating the pre-outburst values.}
    \label{fig:productionrates}
\end{figure*}

Our C$_2$, NH$_2$, and CN production rates are provided in Table~\ref{tab:productionrates} and shown in Figure~\ref{fig:productionrates}, together with the values in the literature. 
The production rates display different trends with time and heliocentric distances. 
We measured a slight decrease of the CN production rate between November 14/15 and December 23 when 2I was at 2.08 au from the Sun pre-perihelion and 2.03 au post-perihelion respectively. 
After that date, we saw a relatively constant CN production rate with heliocentric distance. 
The CN production rate increased for our last two measurements on March 16 and 19. 
Unlike CN, both the NH$_2$ and the C$_2$ production rates decreased from the start of our observations until February 28 (r$_h$ = 2.7 au), but showed an increase similar to CN on March 16 and 19. 
We discuss rate changes relative to 2I's splitting event in \S \ref{sec:postsplit}.

Other authors have previously reported gas production rates for 2I at a similar time to some of our observations. 
As illustrated in Figure \ref{fig:productionrates}, our measurements are consistent with the production rates of C$_2$ and CN in \cite{Lin:2020}.
They are significantly lower than those reported by \cite{Aravind:2021} for 2019 Dec 22 and 2019 Nov 30; the discrepancy could be due to the difference in aperture used to extract the spectrum or the velocity used\footnote{\cite{Aravind:2021} use 1 km/s (K. Aravind, pers. comm.) while we use 0.5 km/s.} in the Haser model. 
If we use the same velocity, our measurements are mostly in agreement with those from \cite{Aravind:2021} at the end of November. 
We also note that the CN band we used is different from the CN band used in most other studies, which could contribute to discrepancies. 
Our rates of NH$_2$ production are several times higher than those of \citet{Prodan:2024} for a similar date. 
The reason for this discrepancy is difficult to identify, but could be attributed to the different fluorescence bands ((0,10,0) vs (0,8,0)), or a difference in aperture used. 
In this work, we used a relatively small aperture of 5,000~km to avoid contamination by stars as 2I was crossing the galactic plane. 
If the radial profile of the species does not match the Haser profile well (which we cannot verify due to low SNR), the aperture size will have a significant impact on the production rates derived. 

\begin{figure}
    \includegraphics[width=1\columnwidth]{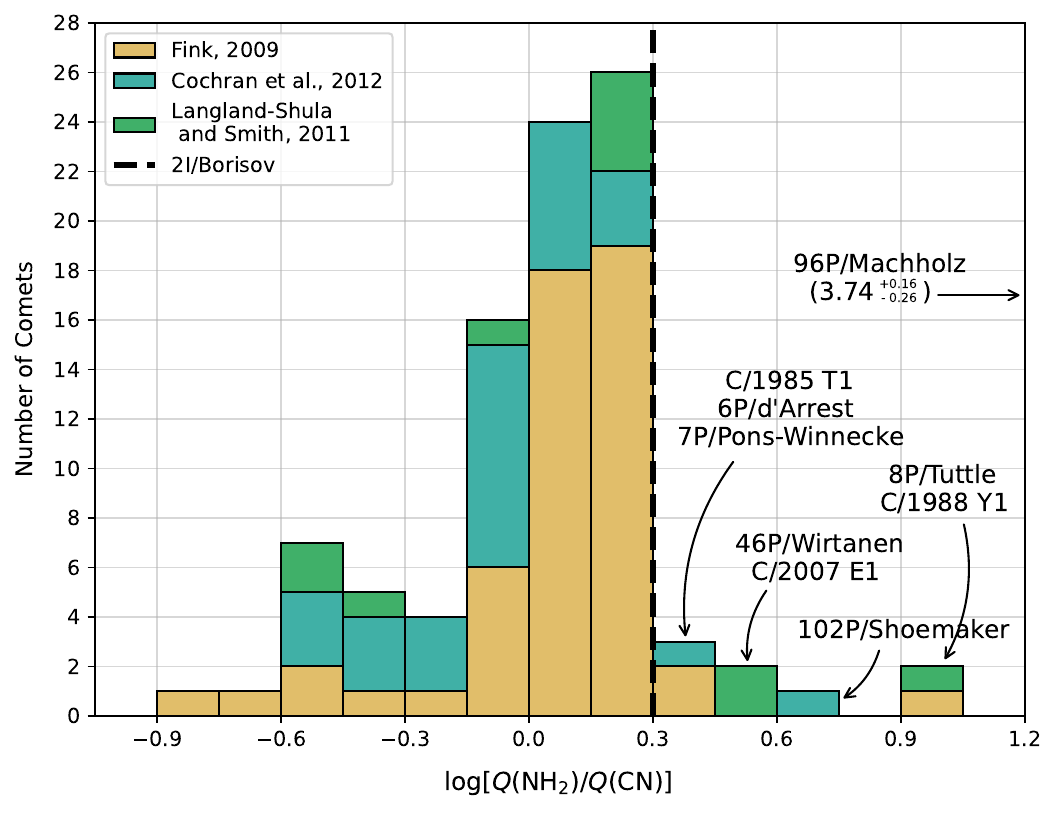}
    \caption{The ratio of NH$_2$ to CN production for Solar System comets in comparison to 2I/Borisov (dashed line). 
    Several other comets from the collections of \citet{Fink:2009}, \citet{Cochran:2012}, and \citet{Langland-Shula:2011} have shown NH$_2$ to CN ratios higher than 2I/Borisov; these are individually labelled. 
    Note the histogram bars per collection are stacked.}
    \label{fig:2IvsSScomets}
\end{figure}

The production rates presented here are slightly different from those reported in \cite{Bannister:2020} using the same data, but consistent within the uncertainty. 
This could be attributed to a combination of factors. 
First, the aperture used is different: 10\arcsec\ in \cite{Bannister:2020}, corresponding to about 15,000 to 17,000~km versus 5,000~km here. 
The telluric absorption was also treated differently. 
In \cite{Bannister:2020} we used the telluric correction from the MUSE pipeline, while here we performed an actual fit of the telluric features using the Molecfit software, leading to a better correction in the redder part of the spectrum where the CN emission is located. 
This is consistent with the largest differences being observed for the CN.
We provide an update to Figure 5 of \citet{Bannister:2020}, placing 2I's NH$_2$ production within the context of Solar System comets (Fig. \ref{fig:2IvsSScomets}).

\section{Discussion}
\label{sec:discussion}

We computed abundance ratios between the different species we observed; these are shown in Figure~\ref{fig:productionrates}. 
We measured a $Q$(C$_2$)/$Q$(CN) ratio between 0.1 and 0.3. 
This confirms that if 2I were a Solar System comet, it would be in the category of carbon-chain depleted comets, which were defined by \cite{Ahearn:1995} as $Q$(C$_2$)/$Q$(CN)$<$ 0.66. %
Additionally, we saw a tentative decreasing trend in the $Q$(C$_2$)/$Q$(CN) with heliocentric distance in our post-perihelion data. 
This would also be consistent with the $Q$(C$_2$) upper limits reported by \citet{Fitzsimmons:2019}, \citet{Lin:2020}, and \citet{de-Leon:2020} for observations in September and October 2019, some of which result in a $Q$(C$_2$)/$Q$(CN) lower than what we measure in November 2019. 

Changes in the $Q$(C$_2$)/$Q$(CN) ratios during passage in the inner Solar System have been seen for some Solar System comets. 
In their study of a group of comets, \cite{Langland-Shula:2011} found a decrease of the ratio with the heliocentric distance. 
This was also reported by \cite{Opitom:2015} for comet C/2016 F6 (Lemmon). 
The origin of the decrease is still unclear, but it has been postulated that it could be related to the model parameters (e.g. scalelengths) used in the Haser model. 
However, other large studies of comet composition have not shown that this trend exists systematically among comets \citep{Cochran:2012,Ahearn:1995,Fink:2009}. 
Thus, the trend we observe for the $Q$(C$_2$)/$Q$(CN) ratio of 2I/Borisov is rare, but not unprecedented among Solar System comets.

We also measured a $Q$(NH$_2$)/$Q$(CN) ratio in the range 0.7-2.1, with an average of 1.7. 
This value is on the high side compared to \citet{Fink:2009}, \citet{Cochran:2012}, and \citet{Langland-Shula:2011}'s samples of Solar System comets: see Figure \ref{fig:2IvsSScomets}. 
This is consistent with our initial assessment in \cite{Bannister:2020}: 2I is relatively rich in NH$_2$ compared to most Solar System comets.

Over the observing period targeted in this work, we saw a decrease in NH$_2$ and C$_2$ abundances that was sharper than that for CN. 
\cite{Bodewits:2020} report a similar effect for H$_2$O and CO over the same time interval, with the H$_2$O abundances decreasing wile CO remains almost constant. 
This behaviour could be indicative of NH$_2$ and C$_2$ being released from a similar source region as H$_2$O, while CN is more correlated with the CO, or that a region with higher CN and CO abundances was illuminated post-perihelion. 
However, given the uncertainties in the abundances due to the faintness of the target, it is impossible to investigate this further.

Morphological features, such as jet-like features, in a comet's coma have been attributed in the past to active areas on the nucleus of comets. 
Changes in the shape or number of visible features have thus been hypothesised to be linked with changes of the illuminated areas on the nucleus. 
Previous studies focusing on measuring gas production rates in the coma of 2I have shown significant changes in relative abundances in its coma between pre- and post-perihelion observations, which they attributed to a seasonal effect \citep{Xing:2020,Aravind:2021}. 
Our morphological observations of the dust coma of 2I, however, did not reveal any evidence of seasonal effect.

The multi-species production and uniform activity of 2I indicates a volatile-rich object.
This is in contrast to 1I/\okina Oumuamua, which appeared devolatilised with a reddened surface \citep[e.g.][]{Bannister:2017,Fitzsimmons:2018}.
We confirm that 2I/Borisov therefore formed beyond ice lines, in the outer part of its system's disk. 
Our data do not further tie down a disk formation location, or constrain the M-dwarf origin suggested by \citet{Cordiner:2020} and \citet{Bodewits:2020}.

The presence of volatiles indicates 2I had minimal processing prior to leaving its parent system. 
This is supported by the CO and polarisation measurements in other data \citep{Bodewits:2020, Cordiner:2020, Bagnulo:2021, Halder2023}, and the dynamics of ISOs in the Galaxy \citep{Forbes:2025}, which predict that ISOs have exceptionally low probabilities for any close stellar encounters after unbinding, i.e. the default assumption should be that they are not thermally heated during galactic travel.

Looking to our own Solar System as an exemplar, the thermal processing of a comet is intrinsically linked to its dynamical history \citep{Gkotsinas:2024}. 
Bodies which migrated to an Oort cloud-like orbit are more likely to be pristine in nature, while bodies ejected from the Solar System during giant planet migration are statistically more processed.
This could indicate that 2I's ejection mechanism was more akin to gently drifting from an Oort cloud, as opposed to early planetary scattering --- despite ongoing unbinding of Oort cloud objects from systems only contributing $\gtrsim$10~\% of the ISO population \citep{PortegiesZwart:2021}.
For a summary of ISO ejection mechanisms, see Sec. 4 of \citet{Fitzsimmons:2023}.

Studying more active ISOs will enable us to understand whether all ISOs are like 2I or if they have a range of activity, shedding light on the ejection mechanisms of ISOs in the Milky Way.
\citet{Hopkins:2025} find that based on its kinematics, 2I could originate from a star of a broad range of ages, all across the Galaxy; our findings remain consistent with this argument.

\subsection{Post-splitting Behaviour}
\label{sec:postsplit}

\begin{figure}[ht!]
\includegraphics[width=\columnwidth]{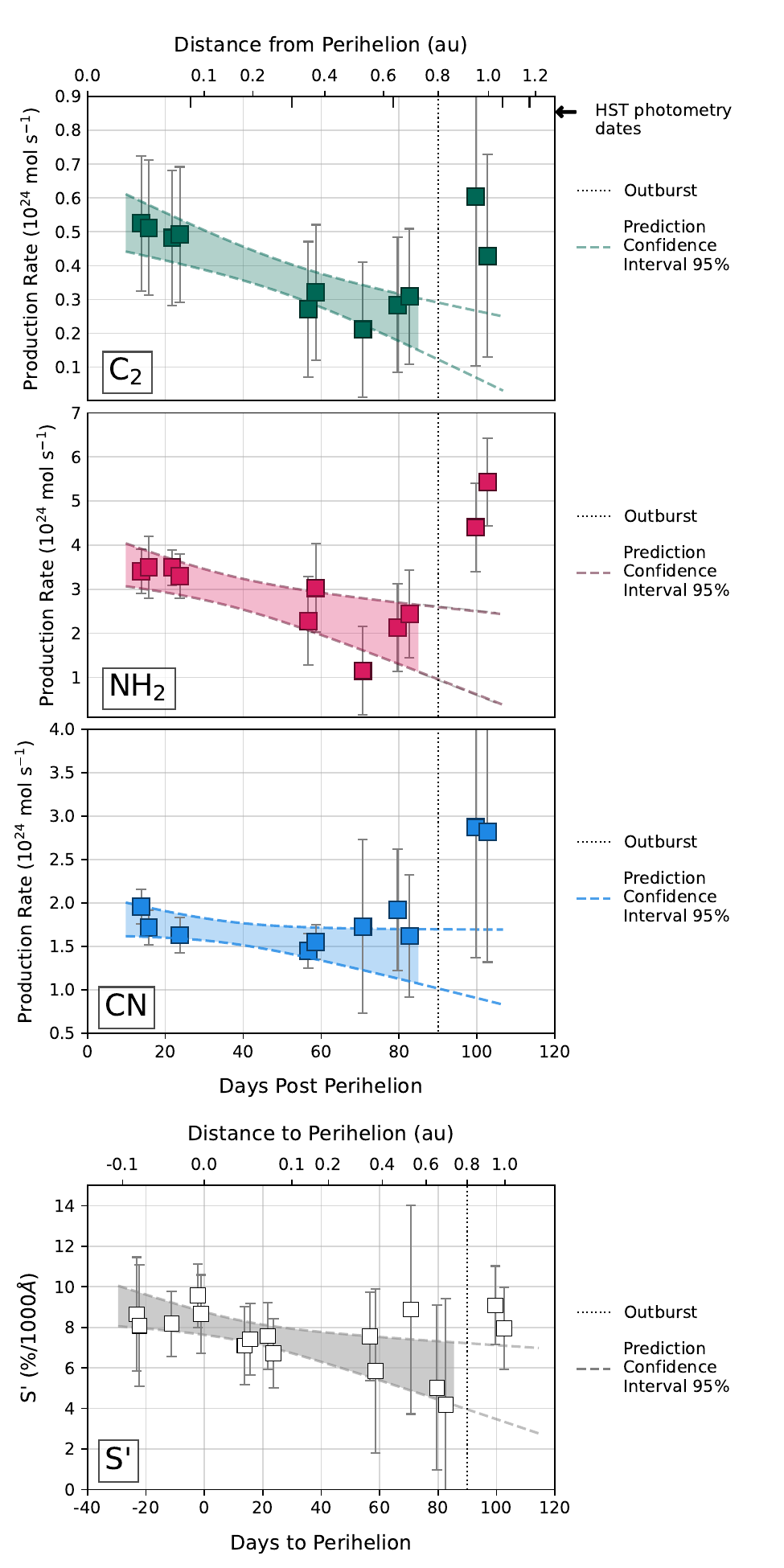}
\caption{The MUSE-observed C$_2$, NH$_2$, \& CN production rates and the dust colour $S'$ measurements. 
The 95\% confidence interval windows for the production rates are predicted on the post-perihelion, pre-outburst values, while the interval for $S'$ is predicted on all pre-outburst values. 
NH$_2$ showed a clear increase in production after the outburst.
C$_2$ and CN also increased but the uncertainties fall within the expected pre-outburst trend. 
$S'$ also shows a minimal increase, i.e. 2I's colour reddened. 
The dates of Hubble Space Telescope (HST) photometry observations from \citet{Jewitt:2020} are indicated by tick marks along the top axis.} 
\label{fig:prodrateschange}
\end{figure}

A brightness increase indicating an outburst of 2I was detected around March 4-9 in ground-based observations \citep{Drahus:2020}.
Spatially separated fragments, and associated brightening from $H$=14.45 to $H$=14.19, were detected in Hubble Space Telescope observations made using the broadband F350LP filter \citep{Jewitt:2020}. %
Our observations and those of \citet{Jewitt:2020} both bracket the split.
Fortuitously, both datasets were acquired by the facilities' respective queues with a very similar observing cadence, differing by only 1-3 days (we indicate the respective cadences in Fig.~\ref{fig:prodrateschange}).
This fragmentation event thus provided a unique opportunity to probe the composition of the interior of 2I's nucleus. 

On fragmenting, 2I produced much more NH$_2$.
The NH$_2$ gas production rate showed a statistically significant ($>95$\% confidence interval) increase above the pre-outburst trends (Figure~\ref{fig:prodrateschange}).
The post-split C$_2$ and CN production rates also showed an increase, though the uncertainties are much larger.
There is a large scatter in the post-perihelion $Q$(NH$_2$)/$Q$(CN) ratios, while $Q$(C$_2$)/$Q$(CN) decreased consistently post-perihelion.
Neither species ratios, however, show a statistical change from the post-perihelion trends after the outburst, so we cannot discern any change in ice composition of 2I due to the outburst.
At the SNR of our measurements, we do not see any change in the gas morphology that could be linked to the outburst.
2I did have a slight increase in the slope of the dust colour (i.e., reddening) after the reported outburst (Fig.~\ref{fig:prodrateschange}).
The measured values lie outside the predictive interval, calculated by extrapolating the pre-outburst values (see Section \ref{sec:methoddustcolour} for details).
This deviation in 2I's dust colour from the expected trend after the outburst is potentially linked to an exposure of larger particles from the freshly exposed surface, and indicative of the top layer of 2I's nucleus being slightly depleted in volatiles compared to the bulk composition.
The dust morphology of 2I (Fig. \ref{dustmapgrid} \& \ref{dustmapenhancedgrid}) does not appear to change with the outburst.

Our observations of 2I's composition do not show any conclusive heterogeneity of the nucleus ice composition. Any changes we see would need to be significantly beyond measurement uncertainties to be certain that our assumptions, for example, the gas profile for the Haser model, are well suited to analysing 2I both before and after the outburst.
A homogeneous nucleus would be similar to the comet 73P/Schwassmann-Wachmann 3, which displayed a constant composition of ices after its fragmentation \citep{DelloRusso2007,Schleicher2011}. 
If 2I did experience some processing, either at its host star or during its interstellar passage, it must have been minimal.

\subsection{Expectations for Observations of Future Interstellar Objects}
\label{sec:future_ISOs}

Upcoming sky surveys such as the Vera Rubin Observatory's Legacy Survey of Space and Time (LSST) and NEOSurveyor are anticipated to provide a substantial sample of additional ISOs to characterise.
Choices will have to be made about target prioritisation.
2I was characterizable with MUSE for a time consistent with the predictions of \citet{Dorsey:2025} (see their Fig. 15). 
We note that LSST would discover a 2I-equivalent object at a greater heliocentric distance than 2I's 3~au, due to the survey's depth being nearly a hundred times greater than 2I's discovery brightness, which would increase the characterisation window from what MUSE achieved here.
However, future ISOs are likely to be found by LSST equally both before and after perihelion \citep{Dorsey:2025}, so not all ISOs will be able to have campaigns throughout perihelion like the one we present here.
In particular, the post-perihelion discoveries only have a median 73-day window for spectroscopic characterisation.
As the LSST sample is predicted to comprise of order tens of ISOs, multi-epoch campaigns on ISOs will continue to be needed for comparative characterisation.
As demonstrated in this work, because of the IFU nature of MUSE, we were able to retrieve a lot of information while 2I was crossing the Galactic plane --- which would not have been possible if we had used a regular long-slit spectrograph. 
As shown in \citet{Hopkins:2025}, future ISOs are likely to spend substantial time on the Galactic plane, which is also the case for the recently announced\footnote{2025-07-02: \url{https://minorplanetcenter.net/mpec/K25/K25N12.html}} third ISO, 3I/ATLAS.
For future rare extended objects that are only visible for a limited time window, using the MUSE instrument would then be an excellent option.

\section{Conclusion}
\label{sec:conclusion}

We used observations with the MUSE IFU to constrain the morphology and composition of 2I/Borisov for an extensive amount of its observing passage.
This forms the longest large-aperture dataset yet acquired for the perihelion passage of an interstellar object, and the only dataset on 2I's composition sampled throughout its entire visible window. 
As the post-perihelion data fell on the Galactic plane, we developed the star-subtraction package \texttt{starkiller} \citep{Ridden-Harper2025}; this significantly improved the usefulness of the data acquired during 2I's recession for assessing the dust in its comae.
From this analysis of dust colour, dust, C$_2$, NH$_2$, \& CN gas species morphology, and the gas species production rates, we are able to see that:
\begin{itemize}
    \item 2I retains volatiles and is uniformly active, indicating minimal processing before it left its parent system.
    \item Considered relative to Solar System comets, 2I is carbon-depleted, but fairly NH$_2$-rich.
    \item On splitting, 2I revealed material that led to a substantial increase in NH$_2$ production.
\end{itemize}

2I displays both similarities and differences from comets in the Solar System.
The common aspects of 2I/Borisov with comets in our own system suggests planetesimal formation processes happen in similar ways across the Galaxy, in both space and time. 
Interstellar objects thus hold both the promise of unveiling the subtleties of protoplanetary disk variation, and affirm the Copernican principle as applied to planetesimal formation across the cosmos.

\begin{acknowledgments}

Based on observations collected at the European Southern Observatory under ESO programmes 103.2033.001--003 and 105.2086.002.
We thank the ESO staff, particularly Henri Boffin, Diego Parraguez, Edmund Christian Herenz, Fuyan Bian, and Israel Blanchard, for their help in the acquisition of these observations.

We would like to thank John Forbes for his helpful advice on statistics. 

MTB appreciates support by the Rutherford Discovery Fellowships from New Zealand Government funding, administered by the Royal Society Te Ap\={a}rangi.

D.Z.S. is supported by an NSF Astronomy and Astrophysics Postdoctoral Fellowship under award AST-2303553. This research award is partially funded by a generous gift of Charles Simonyi to the NSF Division of Astronomical Sciences. The award is made in recognition of significant contributions to Rubin Observatory’s Legacy Survey of Space and Time.

This research made use of the Canadian Advanced Network for Astronomy Research (CANFAR) and the facilities of the Canadian Astronomy Data Centre operated by the National Research Council of Canada with the support of the Canadian Space Agency.
This research has made use of data and/or services provided by the International Astronomical Union's Minor Planet Center.

The views expressed in this article are those of the authors and do not reflect the official policy or position of the U.S. Naval Academy, Department of the Navy, the Department of Defense, or the U.S. Government.

\end{acknowledgments}

\vspace{5mm}
\facilities{VLT(MUSE)}

\software{\texttt{astroplan} \citep{astroplan}, \texttt{astroquery} \citep{astroquery}, Astropy \citep{astropy:2022}, \texttt{jplephem} \citep{PyEphem}, Matplotlib \citep{matplotlib}, MUSE Python Data Analysis Framework (MPDAF) \cite{MPDAFsoftware}, NumPy \citep{numpy}, pandas \citep{pandas:2010, pandas:2020}, Photutils \citep{photutils}, SciPy \citep{scipy}, \texttt{starkiller} \citep{Ridden-Harper2025}, statsmodels \citep{statsmodels}.}

\appendix

\section{Observation Information}
\label{sec:AppendixObservingInfo}

\startlongtable
\begin{deluxetable*}{ccccccccccc}

\tablecaption{VLT/MUSE observations of 2I/Borisov, with geometric relationships between 2I, Earth, and the Sun from JPL Small-Body Database \url{https://ssd.jpl.nasa.gov/horizons/} obtained using \texttt{astroplan} and \texttt{astroquery}.}
\tablehead{\colhead{Observation Date} & \colhead{Altitude$^1$} & \colhead{Airmass$^2$} & \colhead{DIMM Seeing$^3$} & \colhead{$r_h$$^4$} & \colhead{$\Delta$$^5$} & \colhead{Elongation$^6$} & \colhead{$\alpha$$^7$} & \colhead{$\theta_{-\odot}$$^8$} & \colhead{$\theta_{-V}$$^9$} & \colhead{$l^{10}$}\\ 
\colhead{(UTC)}  & \colhead{($\mathrm{{}^{\circ}}$)} & \colhead{} & \colhead{($"$)} & \colhead{($au$)} & \colhead{($au$)} & \colhead{($\mathrm{{}^{\circ}}$)} & \colhead{($\mathrm{{}^{\circ}}$)} & \colhead{($\mathrm{{}^{\circ}}$)} & \colhead{($\mathrm{{}^{\circ}}$)} & \colhead{($km$)}} 
\startdata
2019-Nov-14 08:03:10 & 28.7 & 2.17 & 0.8 & 2.08 & 2.23 & 68.25 & 26.26 & 289.29 & 330.98 & 1617 \\ 
2019-Nov-14 08:19:45 & 32.4 & 1.95 & 0.8 & 2.08 & 2.23 & 68.25 & 26.26 & 289.29 & 330.98 & 1617 \\ 
\rowcolor{lightgray}2019-Nov-14 08:31:36\tablenotemark{a} & 35.0 & 1.82 & 0.72 & 2.08 & 2.23 & 68.25 & 26.26 & 289.29 & 330.98 & 1617 \\ 
\rowcolor{lightgray}2019-Nov-14 08:48:23\tablenotemark{a} & 38.6 & 1.67 & 0.72 & 2.08 & 2.23 & 68.26 & 26.26 & 289.29 & 330.98 & 1617 \\ 
2019-Nov-15 08:02:02 & 29.3 & 2.13 & 0.87 & 2.07 & 2.22 & 68.60 & 26.40 & 289.25 & 330.94 & 1608 \\ 
2019-Nov-15 08:18:33 & 32.9 & 1.92 & 0.68 & 2.07 & 2.22 & 68.61 & 26.40 & 289.25 & 330.93 & 1608 \\ 
\rowcolor{lightgray}2019-Nov-15 08:30:28\tablenotemark{a} & 35.5 & 1.79 & 0.76 & 2.07 & 2.22 & 68.61 & 26.40 & 289.25 & 330.93 & 1608 \\ 
\rowcolor{lightgray}2019-Nov-15 08:53:20\tablenotemark{a} & 40.5 & 1.60 & 0.57 & 2.07 & 2.22 & 68.61 & 26.40 & 289.25 & 330.93 & 1607 \\ 
2019-Nov-26 07:17:24 & 28.0 & 2.16 & 0.43 & 2.02 & 2.09 & 72.41 & 27.69 & 289.11 & 330.02 & 1517 \\ 
2019-Nov-26 07:34:33 & 31.9 & 1.93 & 0.74 & 2.02 & 2.09 & 72.42 & 27.69 & 289.11 & 330.02 & 1517 \\ 
2019-Nov-26 07:46:28 & 34.6 & 1.79 & 0.8 & 2.02 & 2.09 & 72.42 & 27.69 & 289.11 & 330.02 & 1517 \\ 
2019-Nov-26 08:03:33 & 38.5 & 1.64 & 0.74 & 2.02 & 2.09 & 72.43 & 27.70 & 289.11 & 330.02 & 1516 \\ 
2019-Dec-05 07:09:26 & 33.2 & 1.82 & 0.93 & 2.01 & 2.02 & 75.40 & 28.36 & 289.50 & 328.65 & 1462 \\ 
2019-Dec-05 07:26:38 & 37.2 & 1.66 & 1.34 & 2.01 & 2.02 & 75.40 & 28.36 & 289.50 & 328.65 & 1462 \\ 
2019-Dec-05 07:38:31 & 39.9 & 1.57 & 1.17 & 2.01 & 2.02 & 75.40 & 28.36 & 289.50 & 328.65 & 1462 \\ 
2019-Dec-05 07:55:01 & 43.6 & 1.46 & 1.21 & 2.01 & 2.02 & 75.41 & 28.36 & 289.50 & 328.64 & 1461 \\ 
2019-Dec-06 07:37:32 & 40.4 & 1.55 & 1.04 & 2.01 & 2.01 & 75.73 & 28.42 & 289.57 & 328.45 & 1457 \\ 
2019-Dec-06 07:53:57 & 44.1 & 1.45 & 1.61 & 2.01 & 2.01 & 75.73 & 28.42 & 289.57 & 328.45 & 1457 \\ 
2019-Dec-06 08:05:52 & 46.9 & 1.38 & 1.35 & 2.01 & 2.01 & 75.73 & 28.42 & 289.57 & 328.45 & 1456 \\ 
\rowcolor{lightgray}2019-Dec-06 08:25:06\tablenotemark{a} & 51.2 & 1.30 & 1.47 & 2.01 & 2.01 & 75.74 & 28.42 & 289.57 & 328.45 & 1456 \\ 
2019-Dec-21 07:56:19 & 54.8 & 1.21 & 0.49 & 2.03 & 1.94 & 80.39 & 28.60 & 291.52 & 324.50 & 1410 \\ 
2019-Dec-21 08:13:29 & 58.6 & 1.16 & 0.57 & 2.03 & 1.94 & 80.39 & 28.60 & 291.52 & 324.49 & 1410 \\ 
\rowcolor{lightgray}2019-Dec-21 08:25:25\tablenotemark{a} & 61.2 & 1.13 & 0.65 & 2.03 & 1.94 & 80.40 & 28.60 & 291.52 & 324.49 & 1409 \\ 
\rowcolor{lightgray}2019-Dec-21 08:42:26\tablenotemark{a} & 64.9 & 1.10 & 0.51 & 2.03 & 1.94 & 80.40 & 28.60 & 291.53 & 324.49 & 1409 \\ 
2019-Dec-23 05:50:43 & 28.8 & 1.98 & 0.39 & 2.03 & 1.94 & 80.96 & 28.55 & 291.89 & 323.84 & 1407 \\ 
2019-Dec-23 06:07:20 & 32.3 & 1.80 & 0.28 & 2.03 & 1.94 & 80.97 & 28.55 & 291.89 & 323.83 & 1407 \\ 
2019-Dec-23 06:19:16 & 34.9 & 1.69 & 0.44 & 2.03 & 1.94 & 80.97 & 28.55 & 291.89 & 323.83 & 1407 \\ 
2019-Dec-23 06:42:59 & 40.0 & 1.52 & 0.32 & 2.03 & 1.94 & 80.97 & 28.55 & 291.89 & 323.82 & 1407 \\ 
2019-Dec-29 05:18:41 & 26.2 & 2.12 & 0.66 & 2.06 & 1.94 & 82.74 & 28.30 & 293.22 & 321.54 & 1405 \\ 
2019-Dec-29 05:38:15 & 30.2 & 1.89 & 0.71 & 2.06 & 1.94 & 82.74 & 28.29 & 293.23 & 321.53 & 1405 \\ 
2019-Dec-29 05:50:13 & 32.6 & 1.77 & 0.6 & 2.06 & 1.94 & 82.74 & 28.29 & 293.23 & 321.53 & 1405 \\ 
2019-Dec-29 06:09:06 & 36.5 & 1.62 & 0.48 & 2.06 & 1.94 & 82.75 & 28.29 & 293.23 & 321.52 & 1405 \\ 
2019-Dec-31 05:36:51 & 31.1 & 1.83 & 0.41 & 2.07 & 1.94 & 83.33 & 28.18 & 293.74 & 320.68 & 1406 \\ 
2019-Dec-31 05:53:46 & 34.5 & 1.69 & 0.38 & 2.07 & 1.94 & 83.33 & 28.18 & 293.74 & 320.68 & 1406 \\ 
2019-Dec-31 06:05:41 & 36.9 & 1.60 & 0.41 & 2.07 & 1.94 & 83.33 & 28.18 & 293.74 & 320.67 & 1406 \\ 
2019-Dec-31 06:22:04 & 40.2 & 1.49 & 0.32 & 2.07 & 1.94 & 83.34 & 28.17 & 293.74 & 320.67 & 1406 \\ 
2020-Feb-02 04:17:53 & 33.6 & 1.66 & 0.45 & 2.35 & 2.08 & 93.18 & 24.71 & 307.01 & 302.04 & 1512 \\ 
2020-Feb-02 04:34:27 & 35.9 & 1.57 & 0.53 & 2.35 & 2.08 & 93.18 & 24.70 & 307.02 & 302.03 & 1512 \\ 
2020-Feb-02 04:46:27 & 37.6 & 1.52 & 0.64 & 2.35 & 2.08 & 93.18 & 24.70 & 307.03 & 302.02 & 1512 \\ 
2020-Feb-02 05:03:48 & 40.0 & 1.45 & 0.54 & 2.35 & 2.08 & 93.19 & 24.70 & 307.03 & 302.01 & 1512 \\ 
\rowcolor{lightlightgray}2020-Feb-04 03:18:11\tablenotemark{d} & 26.2 & 2.01 & 0.55 & 2.38 & 2.10 & 93.81 & 24.44 & 308.12 & 300.82 & 1522 \\ 
\rowcolor{lightlightgray}2020-Feb-04 03:34:48\tablenotemark{d} & 28.4 & 1.89 & 0.66 & 2.38 & 2.10 & 93.81 & 24.44 & 308.13 & 300.81 & 1522 \\ 
2020-Feb-04 03:46:44 & 30.1 & 1.80 & 0.44 & 2.38 & 2.10 & 93.81 & 24.44 & 308.13 & 300.80 & 1522 \\ 
\rowcolor{lightlightgray}2020-Feb-04 04:05:56\tablenotemark{d} & 32.7 & 1.69 & 0.45 & 2.38 & 2.10 & 93.82 & 24.44 & 308.14 & 300.80 & 1522 \\ 
2020-Feb-16 04:07:13 & 36.9 & 1.53 & 0.6 & 2.53 & 2.20 & 97.83 & 22.77 & 315.86 & 293.87 & 1593 \\ 
2020-Feb-16 04:24:06 & 38.8 & 1.47 & 0.79 & 2.53 & 2.20 & 97.84 & 22.77 & 315.87 & 293.87 & 1593 \\ 
\rowcolor{lightgray}2020-Feb-16 04:35:59\tablenotemark{b} & 40.1 & 1.44 & 0.66 & 2.53 & 2.20 & 97.84 & 22.77 & 315.88 & 293.86 & 1594 \\ 
2020-Feb-16 04:51:50 & 41.8 & 1.39 & 0.47 & 2.53 & 2.20 & 97.84 & 22.77 & 315.88 & 293.86 & 1594 \\ 
2020-Feb-16 05:04:46 & 43.1 & 1.36 & 0.47 & 2.53 & 2.20 & 97.85 & 22.77 & 315.89 & 293.85 & 1594 \\ 
2020-Feb-25 03:28:18 & 35.4 & 1.57 & 0.81 & 2.65 & 2.28 & 101.03 & 21.49 & 322.76 & 289.78 & 1653 \\ 
2020-Feb-25 03:43:47 & 36.9 & 1.52 & 0.61 & 2.65 & 2.28 & 101.04 & 21.49 & 322.77 & 289.78 & 1653 \\ 
2020-Feb-25 03:55:37 & 38.1 & 1.49 & 0.75 & 2.65 & 2.28 & 101.04 & 21.49 & 322.78 & 289.78 & 1653 \\ 
2020-Feb-25 04:11:53 & 39.7 & 1.44 & 0.49 & 2.65 & 2.28 & 101.04 & 21.49 & 322.79 & 289.77 & 1653 \\ 
\rowcolor{lightgray}2020-Feb-28 04:02:19\tablenotemark{c} & 39.4 & 1.45 & 0.72 & 2.70 & 2.31 & 102.14 & 21.06 & 325.32 & 288.69 & 1674 \\ 
2020-Feb-28 04:20:27 & 41.0 & 1.41 & 0.65 & 2.70 & 2.31 & 102.15 & 21.05 & 325.33 & 288.69 & 1674 \\ 
2020-Feb-28 04:32:18 & 42.0 & 1.38 & 0.6 & 2.70 & 2.31 & 102.15 & 21.05 & 325.34 & 288.69 & 1674 \\ 
2020-Feb-28 04:48:45 & 43.3 & 1.35 & 1.22 & 2.70 & 2.31 & 102.15 & 21.05 & 325.35 & 288.68 & 1674 \\ 
\rowcolor{lightlightgray}2020-Mar-16 04:41:49\tablenotemark{d} & 44.3 & 1.33 & 1.23 & 2.96 & 2.48 & 108.59 & 18.61 & 341.85 & 285.55 & 1801 \\ 
2020-Mar-16 04:58:16 & 44.8 & 1.32 & 1.04 & 2.96 & 2.48 & 108.59 & 18.61 & 341.86 & 285.55 & 1801 \\ 
2020-Mar-16 05:10:08 & 45.2 & 1.31 & 1.09 & 2.96 & 2.48 & 108.60 & 18.61 & 341.87 & 285.55 & 1801 \\ 
\rowcolor{lightlightgray}2020-Mar-16 05:28:34\tablenotemark{d} & 45.5 & 1.30 & 1.11 & 2.96 & 2.48 & 108.60 & 18.61 & 341.88 & 285.55 & 1802 \\ 
2020-Mar-19 05:23:42 & 45.3 & 1.31 & 0.61 & 3.00 & 2.52 & 109.74 & 18.19 & 345.14 & 285.50 & 1826 \\ 
2020-Mar-19 05:46:49 & 45.4 & 1.30 & 0.32 & 3.00 & 2.52 & 109.75 & 18.18 & 345.16 & 285.50 & 1826 \\ 
\rowcolor{lightgray}2020-Mar-19 05:58:44\tablenotemark{c} & 45.4 & 1.30 & 0.36 & 3.00 & 2.52 & 109.75 & 18.18 & 345.17 & 285.50 & 1826 \\ 
\rowcolor{lightlightgray}2020-Mar-19 06:14:45\tablenotemark{d} & 45.2 & 1.31 & 0.43 & 3.00 & 2.52 & 109.76 & 18.18 & 345.18 & 285.50 & 1826 \\ 
\enddata

\tablenotetext{a}{Discarded due to high twilight background}
\tablenotetext{b}{Discarded due to thin cloud conditions}
\tablenotetext{c}{Discarded due to 2I passing in front of a star}
\tablenotetext{d}{Omitted from production rates analysis due to 2I passing in front of a star}
\tablecomments{1: Starting altitude, 2: Starting airmass, 3: FWHM seeing value at the beginning of the observation measured by the ASM-DIMM telescope, 4: Heliocentric distance, 5: Geocentric distance, 6: Solar elongation angle, 7: Solar phase angle, 8: Position angle of the anti-Solar direction, 9: Position angle of the negative velocity vector, 10: Physical size of 1" projected to the distance of 2I in km.}
\label{observations_full}
\end{deluxetable*}

\begin{deluxetable}{ccc}
\tablecaption{Spectrophotometric standard stars imaged on each epoch and used in flux calibrations.}
\tablehead{\colhead{Observation Date} & \colhead{Object Name} & \colhead{Airmass}}
\label{tab:standardstars}
\startdata
14 Nov 2019 & LDS749B   &   1.2 \\
15 Nov 2019 & CPD-69 177 &  1.5 \\
26 Nov 2019 & CPD-69 177 & 1.4 \\
05 Dec 2019 & Feige110   &  1.2 \\
05 Dec 2019 & LTT3218    & 1.0 \\
06 Dec 2019 & LTT3218    & 1.0 \\
21 Dec 2019 & Feige110   & 1.3 \\
23 Dec 2019 & CPD-69 177 & 1.4 \\
23 Dec 2019 & GD71       & 2.8 \\
29 Dec 2019 & Feige110    & 1.8 \\
29 Dec 2019 & GD71 H13.90 & 1.4 \\
31 Dec 2019 & Feige110    & 1.4-1.5 \\
02 Feb 2020 & CPD-69 177  & 1.5 \\
04 Feb 2020 & GD71        & 1.3 \\
16 Feb 2020 & GD71        & 1.3 \\
16 Feb 2020 & HD49798     & 1.1 \\
25 Feb 2020 & GD71        & 1.3 \\
28 Feb 2020 & LTT3218     & 1.1 \\
28 Feb 2020 & EG274       & 1.1 \\
16 Mar 2020 & GD71        & 1.4 \\
16 Mar 2020 & EG274       & 1.0 \\
19 Mar 2020 & GD108 & 1.1 \\
\enddata
\end{deluxetable}

\section{Data for Figures}
\label{sec:AppendixDataforFigures}

\begin{deluxetable*}{ccccc}
\tablewidth{\textwidth}
\tablecaption{Literature and MUSE-measured $S'$.}
\tablehead{\colhead{S'} & \colhead{Filters/Method} &\colhead{Range Used} & \colhead{Dates} & \colhead{Source}\\ \colhead{(\%/1000~\AA{})} & \colhead{-} &  \colhead{(nm)} &\colhead{-} &\colhead{-}}
\label{S' table}
\startdata
        12.5 & g' r' & 475 - 630 & 2019-Sep-10 & \cite{Guzik:2020} \\ 
        12.5 & g' r' & 475 - 630 & 2019-Sep-13 & \cite{Guzik:2020} \\ 
        12 $\pm$ 1 & spectroscopic & 400 - 900 & 2019-Sep-13 & \cite{de-Leon:2020} \\ 
        6 & spectroscopic & 800 - 2400 & 2019-Sep-19 & \cite{Yang:2020} \\ 
        19.9 $\pm$ 1.5 & spectroscopic & 390 - 600 & 2019-Sep-20 & \cite{Fitzsimmons:2019} \\ 
        12 $\pm$ 1 & spectroscopic & 400 - 900 & 2019-Sep-24 & \cite{de-Leon:2020} \\ 
        22 $\pm$ 1 & spectroscopic & 390 – 600 & 2019-Sep-24 & \cite{de-Leon:2020} \\ 
        6 & spectroscopic & 800 - 2400 & 2019-Sep-24 & \cite{Yang:2020} \\ 
        4 & B R & 440 - 650 & 2019-Sep-26 & \cite{Jewitt:2019} \\ 
        12 $\pm$ 1 & spectroscopic & 400 - 900 & 2019-Sep-26 & \cite{de-Leon:2020} \\ 
        22 $\pm$ 1 & spectroscopic & 390 – 600 & 2019-Sep-26 & \cite{de-Leon:2020} \\ 
        6 & spectroscopic & 800 - 2400 & 2019-Oct-09 & \cite{Yang:2020} \\ 
        22 $\pm$ 4 & spectroscopic & 445 - 526 & 2019-Oct-10 & \cite{Kareta:2020} \\ 
        11 $\pm$ 3 & spectroscopic & 526 - 713 & 2019-Oct-10 & \cite{Kareta:2020} \\ 
        10.5 & R I & 647 - 797 & 2019-Oct-19 & \cite{Mazzotta-Epifani:2021} \\ 
        19.3 & spectroscopic & 390 - 600 & 2019-Oct-31  & \cite{Lin:2020} \\ 
        9.2 & spectroscopic & 550 - 900 & 2019-Oct-31  & \cite{Lin:2020} \\ 
        19.3 & spectroscopic & 390 - 600 & 2019-Nov-04  & \cite{Lin:2020} \\ 
        9.2 & spectroscopic & 550 - 900 & 2019-Nov-04  & \cite{Lin:2020} \\ 
        8.6 $\pm$ 2.8 & ~ & 528 – 860 & 2019-Nov-14 & This work \\ 
        8.1 $\pm$ 3.0 & ~ & 528 – 860 & 2019-Nov-15 & This work \\ 
        11 $\pm$ 0.9 & photometry & 330 - 1100 & 2019-Nov-24 & \cite{Yang:2021} \\ 
        8.2 $\pm$ 1.6 & ~ & 528 – 860 & 2019-Nov-26 & This work \\ 
        18 $\pm$ 2 & spectroscopy & 400 - 500 & 2019-Nov-27 & \cite{Prodan:2024} \\
        11 $\pm$ 1 & spectroscopy & 570 - 725 & 2019-Nov-27 & \cite{Prodan:2024} \\
        9.9 $\pm$ 1.2 & R I & 640 - 790 & 2019-Nov-30 & \cite{Aravind:2021} \\ 
        13.5 $\pm$ 1.5 & B V & 420 - 550 & 2019-Nov-30 & \cite{Aravind:2021} \\ 
        5 & R I & 647 – 797 & 2019-Dec-02 & \cite{Mazzotta-Epifani:2021} \\ 
        9.6 $\pm$ 1.6 & ~ & 528 – 860 & 2019-Dec-05 & This work \\ 
        8.7 $\pm$ 1.9 & ~ & 528 – 860 & 2019-Dec-06 & This work \\ 
        7.1 $\pm$ 1.9 & ~ & 528 – 860 & 2019-Dec-21 & This work \\ 
        9.9 $\pm$ 1.2 & R I & 640 - 790 & 2019-Dec-22 & \cite{Aravind:2021} \\ 
        13.5 $\pm$ 1.5 & B V & 420 - 550 & 2019-Dec-22 & \cite{Aravind:2021} \\ 
        7.4 $\pm$ 1.8 & ~ & 528 – 860 & 2019-Dec-23 & This work \\ 
        9.9 $\pm$ 1.2 & R I & 640 - 790 & 2019-Dec-24 & \cite{Aravind:2021} \\ 
        13.5 $\pm$ 1.5 & B V & 420 - 550 & 2019-Dec-25 & \cite{Aravind:2021} \\ 
        11.4 $\pm$ 1.0 & photometry & 330 - 1100 & 2019-Dec-26 & \cite{Yang:2021} \\ 
        7.6 $\pm$ 1.6 & ~ & 528 – 860 & 2019-Dec-29 & This work \\ 
        6.7 $\pm$ 1.7 & ~ & 528 – 860 & 2019-Dec-31 & This work \\ 
        7.6 $\pm$ 2.2 & ~ & 528 – 860 & 2020-Feb-02 & This work \\ 
        5.8 $\pm$ 4.0 & ~ & 528 – 860 & 2020-Feb-04 & This work \\ 
        8.9 $\pm$ 5.2 & ~ & 528 – 860 & 2020-Feb-16 & This work \\ 
        9.3 $\pm$ 3 & photometry & 330 - 1100 & 2020-Feb-19 & \cite{Yang:2021} \\ 
        5.0 $\pm$ 4.1 & ~ & 528 – 860 & 2020-Feb-25 & This work \\ 
        4.2 $\pm$ 5.3 & ~ & 528 – 860 & 2020-Feb-28 & This work \\ 
        9.1 $\pm$ 1.9 & ~ & 528 – 860 & 2020-Mar-16 & This work \\ 
        7.9 $\pm$ 2.0 & ~ & 528 – 860 & 2020-Mar-19 & This work \\ 
        10.6 $\pm$ 1.4 & g' i' & 470 - 762 & 2019-Sep to 2020-Jan & \cite{Hui:2020} \\ 
\enddata
\tablecomments{See also \citet{Busarev:2021} for reflectance curves of 2I/Borisov spanning U, B, V, r, \& i.}
\end{deluxetable*}

\section{Gas Emission Maps}
\label{sec:Appendix:GasEmissionMaps}

\begin{figure*}
\includegraphics[width=1\textwidth]{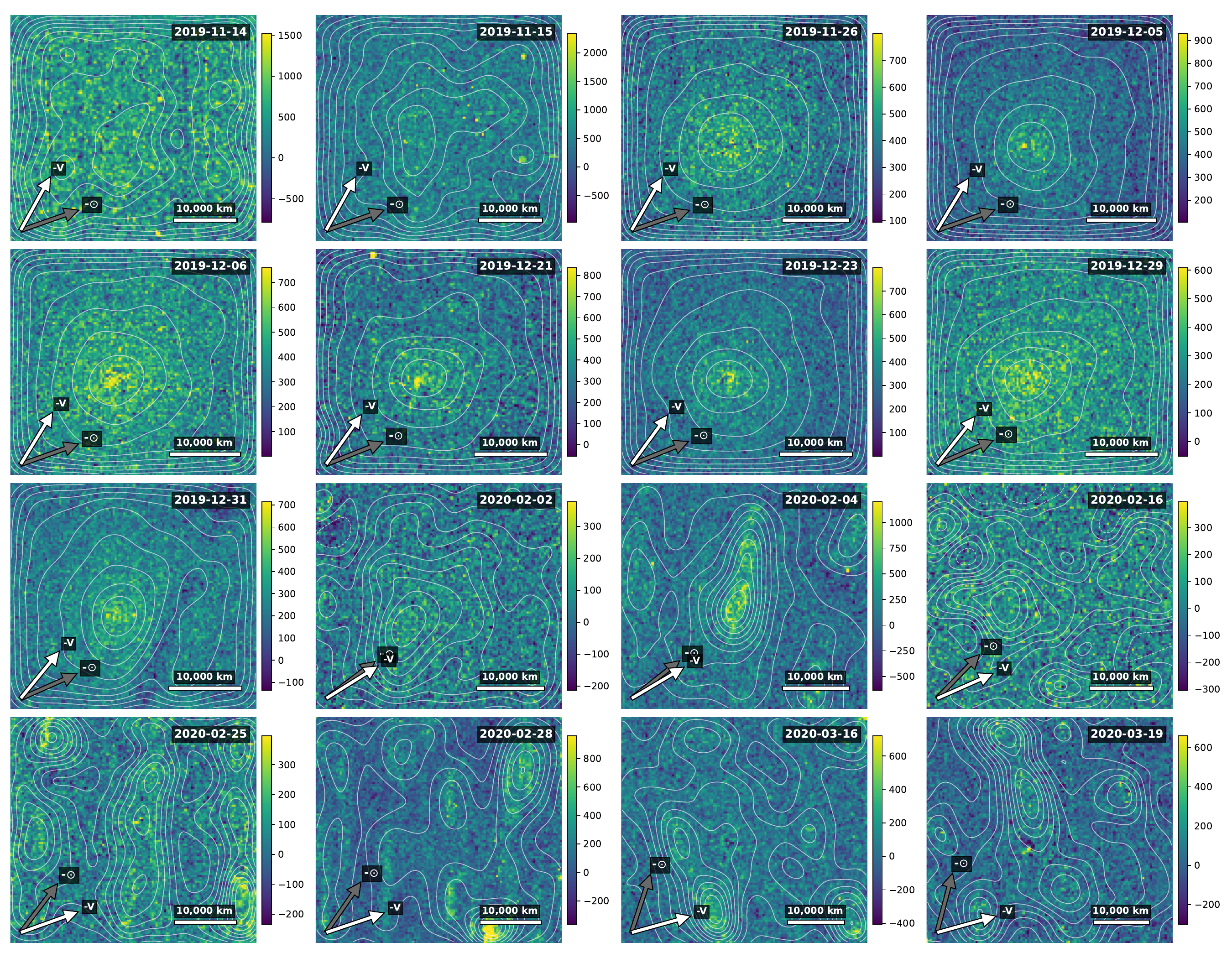}
\caption{The C$_2$ emission maps. Each image is centered on 2I and displayed with a linear stretch over a zscale interval. North is up, east is to the left, and the anti-Solar (-$\odot$) and negative velocity (\textbf{-V}) directions are shown with the respective arrows. The clustering of contours near the image borders is because not all exposures span the entire frame once 2I is in the centre, caused by the dithering in the imaging technique, resulting in less frames contributing to the coadd along the borders.}
\label{C2mapgrid}
\end{figure*}

\begin{figure*}
\includegraphics[width=1\textwidth]{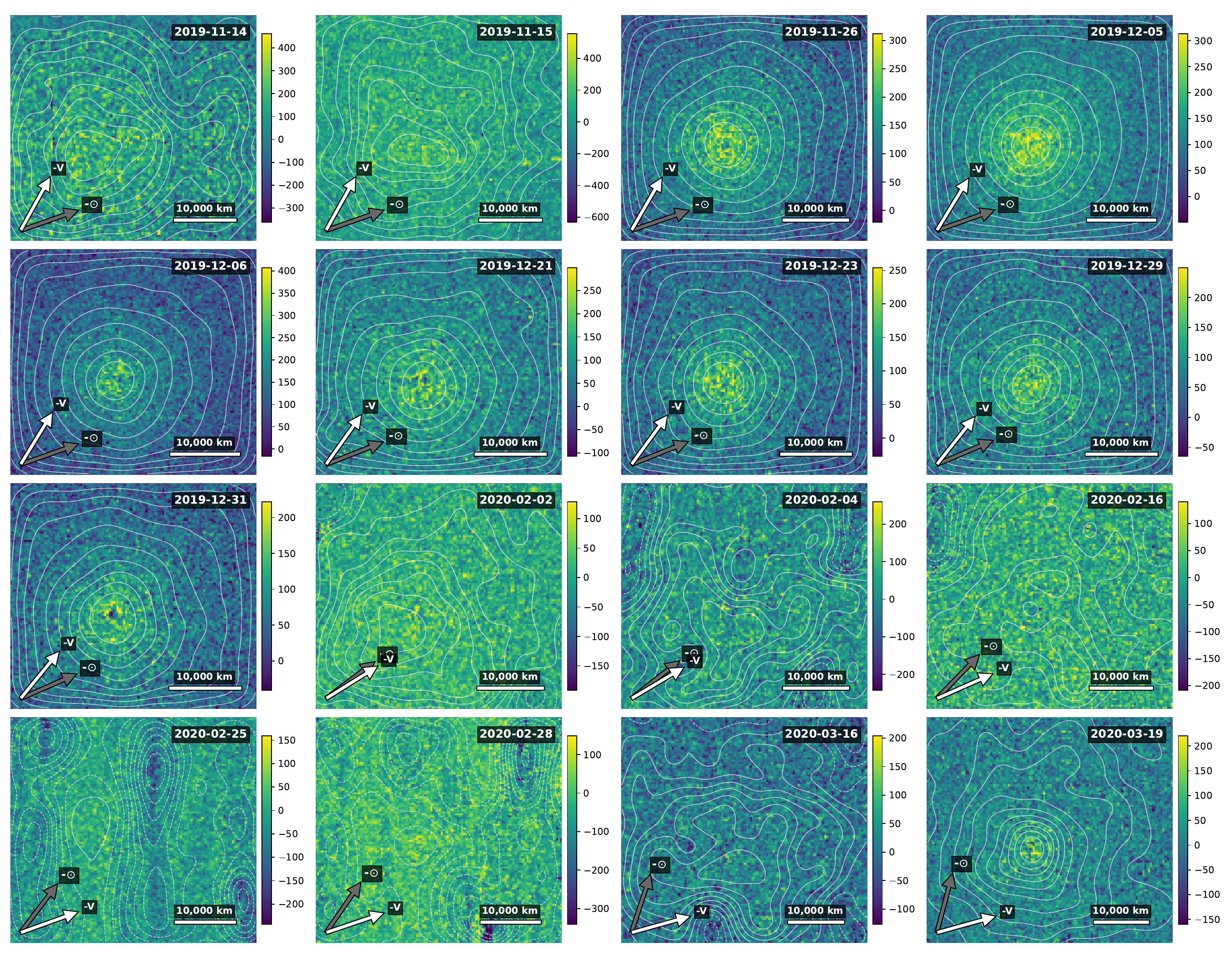}
\caption{The NH$_2$ emission maps. Details are the same as Figure \ref{C2mapgrid}.}
\label{NH2mapgrid}
\end{figure*}

\begin{figure*}
\includegraphics[width=1\textwidth]{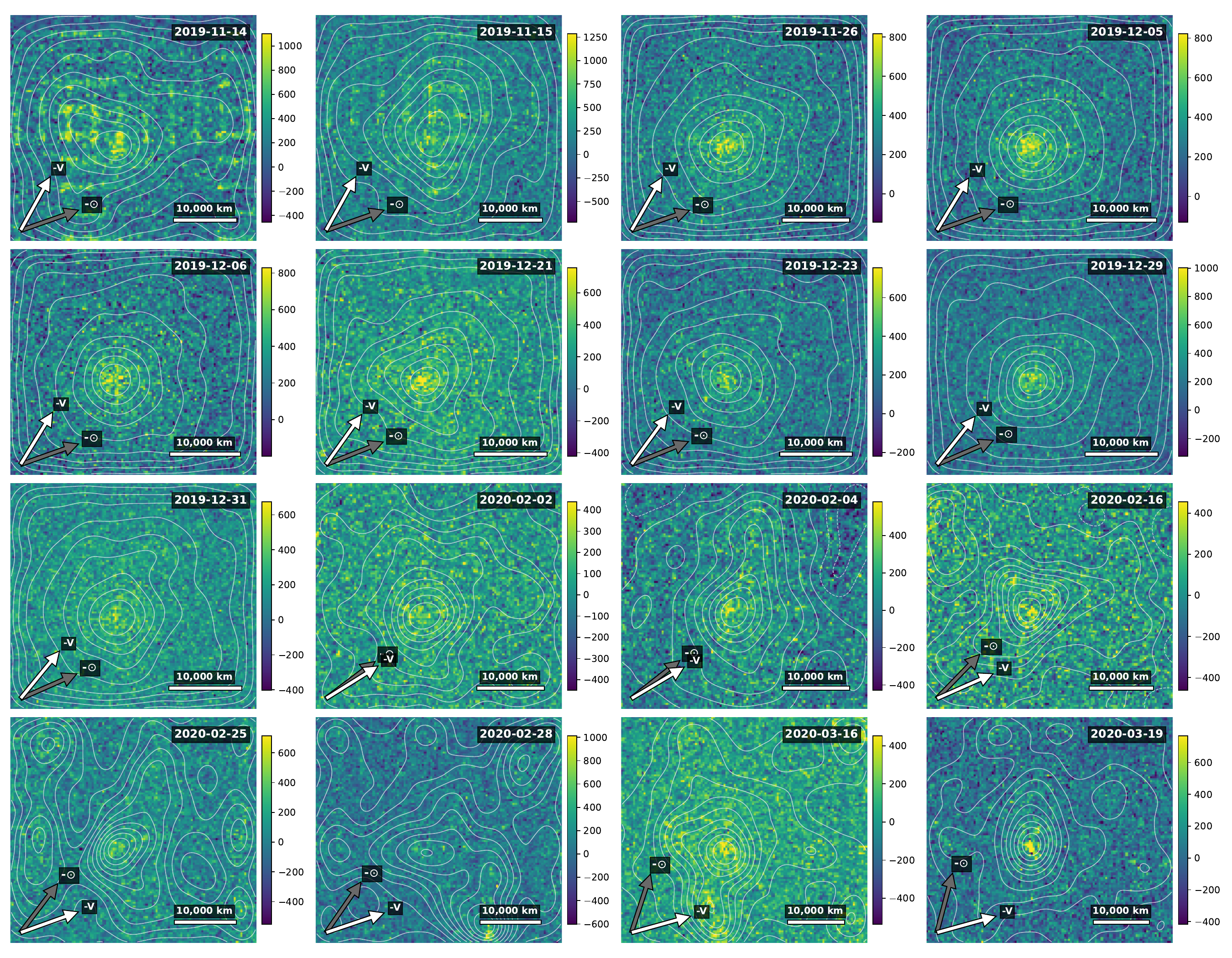}
\caption{The CN emission maps. Details are the same as Figure \ref{C2mapgrid}.}
\label{CNmapgrid}
\end{figure*}

\begin{figure*}
\includegraphics[width=1\textwidth]{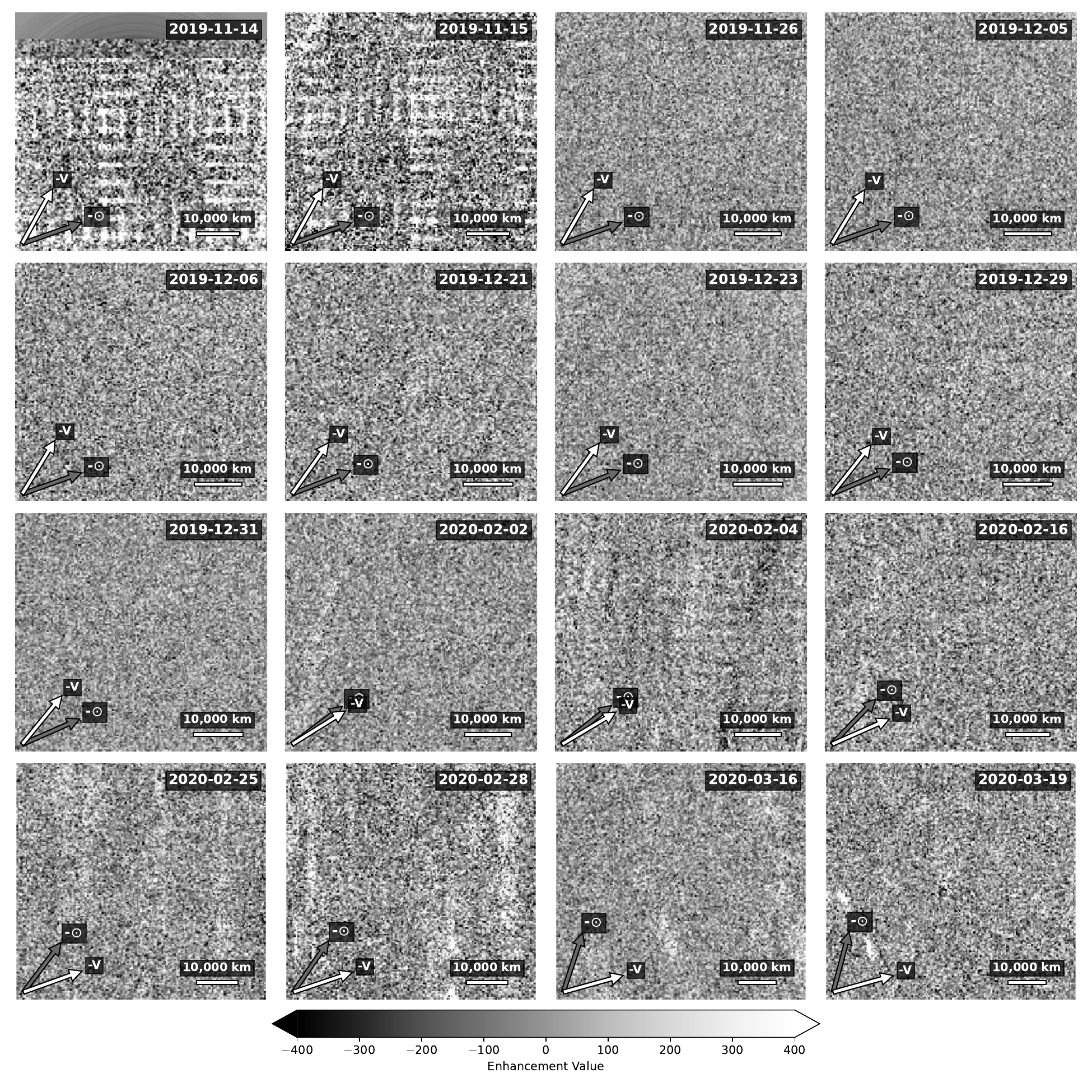}
\caption{The CN emission maps enhanced by subtraction of azimuthal median. Details are the same as Figure \ref{C2mapgrid}. The featureless images indicate uniform CN emission around the optocentre. The criss-cross pattern is an artifact from the IFU borders when less than four exposures were coadded.}
\label{EnhancedCNmapgrid}
\end{figure*}

\newpage

\bibliography{references}{}
\bibliographystyle{aasjournal}

\end{document}